\documentclass[11pt]{llncs}
\usepackage{amssymb,amsmath,enumerate,graphicx}

\usepackage{color}

\newcommand{\red}{\textcolor{red}}

\usepackage{tikz}
\usetikzlibrary{snakes}
\tikzstyle{hyb}=[rectangle,fill=green!50,draw,minimum size=5mm]
\tikzstyle{tre}=[circle,fill=green!50,draw,minimum size=5.5mm]
\tikzstyle{hybnou}=[rectangle,fill=red!50,draw,minimum size=5mm]
\tikzstyle{trenou}=[circle,fill=red!50,draw,minimum size=5.5mm]
\newcommand{\etq}[1]{%
\draw (#1) node {$#1$};
}

\spnewtheorem{algorithm}{Algorithm}{\bfseries}{\itshape}

\newcommand{\pathgr}{\!\rightsquigarrow\!{}}

\newcommand{\LL}{S}
\newcommand{\child}{\mathrm{child}}
\renewcommand{\leq}{\leqslant}
\renewcommand{\geq}{\geqslant}

\renewcommand{\ge}{\geqslant}
\newcommand{\NN}{\mathbb{N}}
\newcommand{\RR}{\mathbb{R}}

\begin{document}

\title{Comparison of Tree-Child Phylogenetic Networks}

\author{Gabriel Cardona\inst{1} \and Francesc Rossell\'o\inst{1} \and
Gabriel Valiente\inst{2} } \authorrunning{G. Cardona et al.}
\institute{Department of Mathematics and Computer Science, University
of the Balearic Islands, E-07122 Palma de Mallorca,
\{\texttt{gabriel.cardona,cesc.rossello}\}\texttt{@uib.es} \and
Algorithms, Bioinformatics, Complexity and Formal Methods Research
Group, Technical University of Catalonia, E-08034 Barcelona,
\texttt{valiente@lsi.upc.edu} }

\maketitle

\begin{abstract}
Phylogenetic networks are a generalization of phylogenetic trees that
allow for the representation of non-treelike evolutionary events, like
recombination, hybridization, or lateral gene transfer.  
While much progress has been made to find practical algorithms for
reconstructing a phylogenetic network from a set of sequences, all
attempts to endorse a class of phylogenetic networks (strictly
extending the class of phylogenetic trees) with a well-founded
distance measure have, to the best of our knowledge, failed so far.
In this paper, we present and study a new meaningful class of
phylogenetic networks, called \emph{tree-child phylogenetic networks},
and we provide an injective representation of these networks as
multisets of vectors of natural numbers, their \emph{path multiplicity
vectors}. We then use this representation to define a distance on this
class that extends the well-known Robinson-Foulds distance for
phylogenetic trees, and to give an alignment method for pairs of
networks in this class. Simple, polynomial algorithms for
reconstructing a tree-child phylogenetic network from its path
multiplicity vectors, for computing the distance between two
tree-child phylogenetic networks, and for aligning a pair of
tree-child phylogenetic networks, are provided. They have been
implemented as a Perl package and a Java applet, and they are available at
the Supplementary Material web page.


\end{abstract}

\section{Introduction}
\label{sec:intro}

Phylogenetic networks have been studied over the last years as a
richer model of the evolutionary history of sets of organisms than
phylogenetic trees, because they take not only mutation events but
also recombination, hybridization, and lateral gene transfer events into account.

The problem of reconstructing a phylogenetic network with the least
possible number of recombination events is
NP-hard~\cite{wang.ea:2001}, and much effort has been devoted to
bounding the number of recombination events needed to explain the
evolutionary history of a set of
sequences~\cite{bafna.ea:2004,myers.ea:2003,song.ea:2005}.  On the
other hand, much progress has been made to find practical algorithms
for reconstructing a phylogenetic network from a set of
sequences~\cite{gusfield.ea:fine.structure:2004,gusfield.ea:galled.trees:2004,moret.ea:2004,nakhleh.ea:2003,nakhleh.ea:2005,song.ea:2005}.

Since different reconstruction methods applied to the same sequences, or
a single method applied to different sequences, may yield different
phylogenetic networks for a given set of species, a sound measure to compare
phylogenetic networks becomes necessary~\cite{nakhleh.ea:03psb}.  The
comparison of phylogenetic networks is also needed in the
assessment of phylogenetic reconstruction methods
\cite{moret:alenex05}, and it will be required to perform queries on
the future databases of phylogenetic networks~\cite{page:05}.

Many metrics for the comparison of phylogenetic trees are known, including the
Robin\-son-Foulds metric~\cite{robinson.foulds:mb81}, the
nearest-neighbor interchange metric~\cite{waterman.smith:1978}, the
subtree transfer distance~\cite{allen.steel:2001}, the quartet
metric~\cite{estabrook.ea:1985}, and the metric from the nodal
distance algorithm~\cite{bluis.ea:2003}.  But, to our knowledge, only
one metric (up to small variations) for phylogenetic networks  has been
proposed so far. It is the so-called \emph{error}, or
\emph{tripartition}, \emph{metric}, developed by Moret, Nakhleh,
Warnow and collaborators in a series of papers devoted to the study
of reconstructibility of phylogenetic networks~\cite{nakhleh.ea:TR0326,%
nakhleh.ea:tutpsb04,nakhleh.ea:TR0209,moret.ea:2004,%
nakhleh:phd04,nakhleh.ea:TR0406,nakhleh.ea:03psb}, and which we recall
in \S\ref{subsec:metr} below.  Unfortunately, it turns out that, even
in its strongest form \cite{moret.ea:2004}, this error metric never
distinguishes all pairs of phylogenetic networks that, according to
its authors, are distinguishable: see \cite{cardona.ea:07a} for
a discussion of the error metric's downsides.

The main goal of this paper is to introduce a metric on a restricted,
but meaningful, class of phylogenetic networks: the \emph{tree-child}
phylogenetic networks.  These are the phylogenetic networks where
every non-extant species has some descendant through mutation.  This
is a slightly more restricted class of phylogenetic networks than the
\emph{tree-sibling} ones (see \S \ref{subsec:phnet}) where one of the
versions of the error metric was defined.  Tree-child phylogenetic
networks include galled
trees~\cite{gusfield.ea:fine.structure:2004,gusfield.ea:galled.trees:2004}
as a particular case, and they have been recently proposed by S. J
Wilson as the class where meaningful phylogenetic networks should be
searched \cite{wilson:07}.

We prove that each tree-child phylogenetic network with $n$ leaves can
be singled out, up to isomorphisms, among all tree-child phylogenetic
networks with $n$ leaves by means of a finite multisubset of $\NN^n$.
This multiset of vectors consists of the \emph{path multiplicity
vectors}, or \emph{$\mu$-vectors} for short, $\mu(v)$ of all nodes $v$
of the network: for every node $v$, $\mu(v)$ is the vector listing the
number of paths from $v$ to each one of he leaves of the network.  We
present a simple polynomial time algorithm for reconstructing a
tree-child phylogenetic network from the knowledge of this multiset.

This injective representation of tree-child phylogenetic networks as
multisubsets of vectors of natural numbers allows us to define a
metric on any class of tree-child phylogenetic networks with the same
leaves as simply the symmetric difference of the path multiplicity
vectors multisets.  This metric, which we call \emph{$\mu$-distance},
extends to tree-child phylogenetic networks the Robinson-Foulds metric
for phylogenetic trees, and it satisfies the axioms of distances,
including the separation axiom (non-isomorphic phylogenetic networks
are at non-zero distance) and the triangle inequality.

The properties of the path multiplicity representation of tree-child
phylogenetic networks allow us also to define an alignment method for
them.  Our algorithm outputs an injective matching from the network
with less nodes into the other network that minimizes in some specific
sense the difference between the $\mu$-vectors of the matched nodes.
Although several alignment methods for phylogenetic trees are
known~\cite{munzner.ea:03,nye.ea:2006,page:95}, this is to our
knowledge the first one that can be applied to a larger class of
phylogenetic networks.

We have implemented our algorithms to recover a tree-child
phylogenetic network from its data multiplicity representation and to
compute the $\mu$-distance, together with other related algorithms
(like for instance the systematic and efficient generation of all
tree-child phylogenetic networks with a given number of leaves), in a
Perl package which is available at the Supplementary Material web
page.  We have also implemented our alignment method as a Java applet
which can be run interactively at the aforementioned web page.

The plan of the rest of the paper is as follows.  In
Section~\ref{sec:prel} we gather some preliminary material: we fix
some notations and conventions on directed acyclic graphs, and we
recall several notions related to phylogenetic trees and networks, the
Robinson-Foulds metric for the former and the tripartition metric for
the latter.  In Section~\ref{sec:TC} we introduce the tree-child
phylogenetic networks and we study some of their basic properties.  In
Section~\ref{sec:mu-representation} we introduce the path multiplicity
representation of networks and we prove that it singles out tree-child
phylogenetic networks up to isomorphism. Then, in Section~\ref{sec:mu-distance}
we define and study the $\mu$-distance for  tree-child
phylogenetic networks with the same number of leaves, and in
Section~\ref{sec:mu-align} we present our alignment method. The paper 
ends with a short Conclusion section.

\section{Preliminaries}
\label{sec:prel}

\subsection{DAGs}
\label{subsec:dags}

Let $N=(V,E)$ be a directed acyclic graph (DAG). 
We denote by $d_i(u)$ and $d_o(u)$ the in-degree and out-degree,
respectively, of a node $u\in V$. 

A node $v\in V$ is a \emph{leaf} if $d_o(v)=0$, and \emph{internal} if
$d_o(v)>0$; a \emph{root} if $d_i(v)=0$; a \emph{tree node} if
$d_{i}(v)\leq 1$, and a \emph{hybrid node} if $d_i(v)>1$.  We denote
by $V_L$, $V_{T}$, and $V_H$ the sets of leaves, of tree nodes, and of
hybrid nodes of $N$, respectively.  A DAG is said to be \emph{rooted}
when it has only one root.

Given an arc $(u,v)\in E$, we call the node $u$ its \emph{tail} and
the node $v$ its \emph{head}.  An arc $(u,v)\in E$ is a \emph{tree
arc} if $v$ is a tree node, and a \emph{hybridization arc} if $v$ is hybrid.
We denote by $E_{T}$ and $E_{N}$ the sets of tree arcs and of hybridization
arcs, respectively.

A node $v\in V$ is a \emph{child} of $u\in V$ if $(u,v)\in V$; we also
say that $u$ is a \emph{parent} of $v$.  For every node $u\in V$, let
$\child(u)$ denote the set of its children.  All children of the same
node are said to be \emph{siblings} of each other.  The \emph{tree
children} of a node $u$ are its children that are tree nodes.

A DAG is \emph{binary} when all its internal tree nodes have
out-degree 2 and all its hybrid nodes have in-degree 2 and out-degree
1.

Let $S$ be any finite set of \emph{labels}.  We say that the DAG $N$
is \emph{labeled in $S$}, or that it is an \emph{$S$-DAG}, for short,
when its leaves are bijectively labeled by elements of $S$.  Two DAGs
$N,N'$ labeled in $S$ are \emph{isomorphic}, in symbols $N\cong N'$,
when they are isomorphic as directed graphs and the isomorphism
preserves the leaves' labels.

In this paper we shall always assume, usually without any further
notice, that the DAGs appearing in it are labeled in some set $S$, and
we shall always identify, usually without any further notice either,
each leaf of a DAG with its label in $S$.

A \emph{path} in $N$ is a sequence of nodes $(v_0,v_1,\dots,v_k)$ such
that $(v_{i-1},v_{i})\in E$ for all $i=1,\dots,k$.  We say that such a
path \emph{starts} in $v_0$, \emph{passes through} $v_1,\dots,v_{k-1}$
and \emph{ends} in $v_k$; consistently, we call $v_{0}$ the
\emph{origin} of the path, $v_{1},\ldots,v_{k-1}$ its
\emph{intermediate} nodes, and $v_{k}$ its \emph{end}.  The
\emph{position} of the node $v_{i}$ in the path $(v_0,v_1,\dots,v_k)$ is
$i+1$.  The \emph{length} of the path $(v_0,v_1,\dots,v_k)$ is $k$,
and it is \emph{non-trivial} if $k\ge 1$: a \emph{trivial} path is,
then, simply a node.  We denote by $u\pathgr v$ any path with origin
$u$ and end $v$.

The \emph{height} of a node is the length of a longest path starting
in the node and ending in a leaf.

We shall say that a path $u\pathgr v$ is \emph{contained} in, or that
it is a \emph{subpath} of, a path $u'\pathgr v'$ when there exist
paths $u'\pathgr u$ and $v\pathgr v'$ such that the path $u'\pathgr
v'$ is the concatenation of the paths $u'\pathgr u$, $u\pathgr v$, and
$v\pathgr v'$.

A path is \emph{elementary} when its origin has out-degree 1 and  all its intermediate nodes have in and
out-degree 1.

The relation $\geq$ on $V$ defined by
$$
u\geq v \iff \mbox{there exists a path $u\pathgr v$}
$$
is a partial order, called the \emph{path ordering} on $N$.  Whenever
$u\geq v$, we shall say that $v$ is a \emph{descendant} of $u$ and
also that $u$ is an \emph{ancestor} of $v$.  For every node $u\in V$,
we shall denote by $C(u)$ the set of all its descendants, and by
$C_{L}(u)$ the set of leaves that are descendants of $u$: we call
$C_{L}(u)$ the \emph{cluster} of $u$.

A node $v$ of $N$ is a \emph{strict descendant} of a node $u$ if it is
a descendant of it, and every path from a root of $N$ to $v$ contains
the node $u$: in particular, we understand every node as a strict
descendant of itself.  For every node $u\in V$, we shall denote by
$A(u)$ the set of all its strict descendants, and by $A_{L}(u)$ the
set of leaves that are strict descendants of $u$: we call $A_{L}(u)$
the \emph{strict cluster} of $u$.

A \emph{tree path} is a non-trivial path such that its end and all its
intermediate nodes are tree nodes.  A node $v$ is a \emph{tree
descendant} of a node $u$ when there exists a tree path from $u$ to
$v$.  For every node $u\in V$, we shall denote by $T(u)$ the set of
all its tree descendants, and by $T_{L}(u)$ the set of leaves that are
tree descendants of $u$: we call $T_{L}(u)$ the \emph{tree cluster} of
$u$.

We recall from \cite{cardona.ea:07a} the following two easy results,
which will be used several times in the next sections.
%


\begin{lemma}\label{lem:unicity-path}
Let $u\pathgr v$ be a tree path.  Then, for every other path $w\pathgr v$
ending in $v$, it is either contained in $u\pathgr v$ or it contains
$u\pathgr v$.
\end{lemma}



\begin{corollary}\label{cor:unicity-path}
If $v\in T(u)$, then $v\in A(u)$ and the path
$u\pathgr v$ is unique.
\end{corollary}
%

\subsection{The Robinson-Foulds metric on phylogenetic trees}
\label{subsec:RF}

A \emph{phylogenetic tree} on a set $S$ of taxa is a rooted tree
without out-degree 1 nodes with its leaves labeled bijectively in $S$,
i.e., a rooted $S$-DAG with neither hybrid nodes nor out-degree 1
nodes.

Every arc $e=(u,v)$ of a phylogenetic tree $T=(V,E)$ on $S$ defines a
\emph{bipartition} of $S$ 
$$
\pi(e)=(C_L(v),S\setminus C_L(v)).
$$
Let $\pi(T)$ denote the set of all these bipartitions:
$$
\pi(T)=\{\pi(e)\mid e\in E\}.
$$

The \emph{Robinson-Foulds metric} \cite{robinson.foulds:mb81} 
between two phylogenetic
trees $T$ and $T'$ on the same set $S$ of taxa is defined as
$$
d_{RF}(T,T')
=|\pi(T)\bigtriangleup \pi(T')|,
$$
where $\bigtriangleup$ denotes the symmetric difference of sets.

The Robinson-Foulds metric is a true distance for phylogenetic trees, in
the sense that it satisfies the axioms of distances up to isomorphisms:
for every phylogenetic trees $T,T',T''$ on the same set $S$ of taxa,
\begin{enumerate}[(a)]
\item \emph{Non-negativity}: $d_{RF}(T,T')\geq 0$
\item \emph{Separation}: $d_{RF}(T,T')=0$ if and only if $T\cong T'$
\item \emph{Symmetry}: $d_{RF}(T,T')=d_{RF}(T',T)$
\item \emph{Triangle inequality}: $d_{RF}(T,T')\leq d_{RF}(T,T'')+d_{RF}(T'',T')$
\end{enumerate}

\subsection{Phylogenetic networks}
\label{subsec:phnet}

A natural model for describing an evolutionary history is a directed
acyclic graph (DAG for short) whose arcs represent the relation
parent-child.  Such a DAG will satisfy some specific features
depending on the nature and properties of this relation.  For
instance, if we assume the existence of a common ancestor of all
individuals under consideration, then the DAG will be \emph{rooted}: it will have only one
root.  If, moreover, the evolutionary history to be described is
driven only by mutation events, and hence every individual has only
one parent, then the DAG will be a tree.  In this line of thought, a
phylogenetic network is defined formally as a rooted DAG with some specific features that are suited to model evolution
under mutation and recombination, but the exact definition varies
from paper to paper: see, for instance,
\cite{bandelt:94,huson:tutgcb06,huson:07,huson.ea:06,nakhleh.ea:tutpsb04,semple:07,strimmer.ea:2000,strimmer.ea:2001}.

For instance, Moret, Nakhleh, Warnow and collaborators have proposed several
slightly different definitions of phylogenetic 
networks~\cite{nakhleh.ea:TR0326,nakhleh.ea:tutpsb04,nakhleh.ea:TR0209,moret.ea:2004,%
nakhleh:phd04,nakhleh.ea:TR0406}.  To recall one of them, in
\cite{nakhleh.ea:TR0326} a \emph{model phylogenetic network} on a set
$S$ of taxa is defined as a rooted $S$-DAG $N$ satisfying the
following conditions:
\begin{enumerate} 
\item[(1.1)] The root and all internal tree nodes have out-degree 2.  All
hybrid nodes have out-degree 1, and they can only have in-degree 2
(allo-polyploid hybrid nodes) or 1 (auto-polyploid hybrid
nodes).

\item[(1.2)] The child of a hybrid node is always a tree node.

\item[(1.3)] \emph{Time consistency:} If $x,y$ are two nodes for which
there exists a sequence of nodes $(v_0,v_{1},\ldots,v_{k})$ with
$v_0=x$ and $v_{k}=y$ such that:
\begin{itemize}
    \item for every $i=0,\ldots,k-1$, either $(v_{i},v_{i+1})$ is an
    arc of $N$, or $(v_{i+1},v_{i})$ is a hybridization arc of $N$,

\item at least one pair $(v_{i},v_{i+1})$ is a tree arc of $N$,
\end{itemize}
then $x$ and $y$ cannot have a hybrid child in common.
\end{enumerate}
(This time compatibility condition (1.3) is equivalent to the
existence of a \emph{temporal representation} of the network
\cite{baroni.ea:sb06,maddison:97}: an assignation of times to the nodes of the
network  that strictly increases on tree arcs and so that the
parents of each hybrid node coexist in time.  See \cite[Thm.~3]{baroni.ea:sb06}
or \cite[Prop.~1]{cardona.ea:07a} for a proof of
this equivalence.)

On the other hand, these authors define in \textsl{loc.  cit.} a
\emph{reconstructible phylogenetic network} as a rooted $S$-DAG where
the previous conditions are relaxed as follows: tree nodes can have
any out-degree greater than 1; hybrid nodes can have any in-degree
greater than 1 and any out-degree greater than 0; hybrid nodes can
have hybrid children; and the time consistency need not hold any
longer.  So, reconstructible phylogenetic networks in this sense are
simply rooted DAGs with neither out-degree 1 tree nodes nor hybrid
leaves.  These model and reconstructible phylogenetic networks are
used, for instance, in \cite{nakhleh.ea:03psb}.

A generalization of reconstructible phylogenetic networks are the
\emph{hybrid phylogenies} of \cite{baroni.ea:ac04}: rooted $S$-DAGs
without out-degree 1 tree nodes.  But although out-degree 1 tree nodes
cannot be reconstructed, they can be useful both from the biological
point of view, to include auto-polyploidy in the model, as well as
from the formal point of view, to restore time compatibility and the
impossibility of successive hybridizations in reconstructed
phylogenetic networks \cite[Fig.  13]{moret.ea:2004}.

In papers on phylogenetic networks it is usual to impose extra
assumptions to the structure of the network, in order to narrow the
output space of reconstruction algorithms or to guarantee certain
desired properties.  For instance,  Nakhleh imposes in his PhD Thesis
\cite{nakhleh:phd04} the \emph{tree-sibling}\footnote{Nakhleh uses the
term \emph{class I} to refer to these networks, but for consistency
with the notations we introduce in the next section, we have renamed
them here.} condition to the phylogenetic networks defined above:
every hybrid node must have at least one sibling that is a tree node.
Although this condition is imposed therein to try to guarantee that
the error metric considered in that work satisfies the separation
axiom of distances (see the next subsection), it has also appeared
under a different characterization in some papers devoted to
phylogenetic network reconstruction algorithms
\cite{nakhleh.ea:bioinfo06,nakhleh.ea:bioinfo07}.  Indeed, the
phylogenetic networks considered in these papers are obtained by
adding hybridization arcs to a phylogenetic tree by repeating the following
procedure:
\begin{enumerate}[1.]
 \item choose pairs of arcs $(u_1,v_1)$ and
$(u_2,v_2)$ in the tree; 
\item split the first into $(u_1,w_1)$ and
$(w_1,v_1)$, with $w_1$ a new (tree) node; 
\item split the second one into
$(u_2,w_2)$ and $(w_2,v_2)$, with $w_2$ a new (hybrid) node; 
\item 
add a new arc $(w_1,w_2)$.
\end{enumerate}
It is not difficult to prove that the phylogenetic networks
obtained in this way are tree-sibling, and that the binary tree-sibling phylogenetic networks are exactly those obtained by applying this procedure to binary phylogenetic trees.

An even stronger condition is the one imposed on \emph{galled
trees}~\cite{gusfield.ea:fine.structure:2004,gusfield.ea:galled.trees:2004,wang.ea:2001}:
no tree node has out-degree 1, all hybrid nodes have in-degree 2, and
no arc belongs to two recombination cycles.  Here, by a
\emph{recombination cycle} we mean a pair of two paths with the same
origin and end and no intermediate node in common.  In the
aforementioned papers these galled trees need not satisfy the time
compatibility condition, but in other works they are imposed to
satisfy it \cite{nakhleh:phd04,nakhleh.ea:TR0406,nakhleh.ea:2005}.

\subsection{Previous work on metrics for phylogenetic networks}
\label{subsec:metr}

While many metrics for phylogenetic trees have been introduced and
implemented in the literature (see, for instance,
\cite{dasgupta.ea:98,puigbo.ea:2007} and the references therein), to
our knowledge the only similarity measures for phylogenetic networks
proposed so far are due to Moret, Nakhleh, Warnow and collaborators in
the series of papers quoted in the last subsection, where they are
applied in the assessment of phylogenetic network reconstruction
algorithms.  We briefly recall these measures in this subsection.

The \emph{error}, or \emph{tripartition}, \emph{metric} is a natural
generalization to networks of the Robinson-Foulds metric for
phylogenetic trees recalled in \S \ref{subsec:RF}.  The basis of this
method is the representation of a network by means of the
tripartitions associated to its arcs.  For each arc $e=(u,v)$ of a DAG
$N$ labeled in $S$, the \emph{tripartition} of $S$ associated to $e$
is
$$
\theta (e)=(A_{L}(v), C_{L}(v)\setminus A_{L}(v), S\setminus C_{L}(v)),
$$
where moreover each leaf $s$ in $A_L(v)$ and $C_{L}(v)\setminus A_{L}(v)$ is weighted with
the greatest number of hybrid nodes contained in a path from $v$ to $s$
(including $v$ and $s$ themselves).\footnote{Actually, Moret, Nakhleh,
Warnow et al consider also other variants of this definition,
 weighting only the non-strict descendant leaves or not
weighting any leaf, but for the sake of brevity and generality we only recall here
the most general version.} Let $\theta(N)$ denote the set of all these
tripartitions of arcs of $N$.

In some of the aforementioned papers, the authors enrich these
tripartitions with an extra piece of information.  Namely, they
define the \emph{reticulation scenario} $RS(v)$ of a hybrid node $v$
with parents $u_{1},u_{2}$ as the set of clusters of its parents:
$$
RS(v)=\{C_{L}(u_{1}),C_{L}(u_{2})\}.
$$
Then, the \emph{enriched tripartition} $\Psi(e)$ associated to an arc
$e$ is defined as $\theta (e)$ if $e$ is a tree arc, and as the pair $(\theta
(e),RS(v))$ if $e$ is a hybridization arc with head $v$.
Let $\Psi(N)$ denote the set of all these enriched tripartitions.

For every $\Upsilon=\theta,\Psi$, the \emph{error}, or
\emph{tripartition}, \emph{metric relative to $\Upsilon$} between two
DAGs $N_{1}=(V_{1},E_{1})$ and $N_{2}=(V_{2},E_{2})$ labeled in the
same set $S$ is defined by these authors as
$$
m_{\Upsilon}(N_{1},N_{2})=\frac{1}{2}\Bigl(\frac{| \Upsilon(N_{1})\setminus
\Upsilon(N_{2})|}{|E_{1}|}+\frac{| \Upsilon(N_{2})\setminus
\Upsilon(N_{1})|}{|E_{2}|}\Bigr).
$$
Unfortunately, and despite the word `metric', this formula does not
satisfy the separation axiom on any of the subclasses of phylogenetic
networks where it is claimed to do so by the authors, and hence it
does not define a distance on them: for instance, $m_{\Psi}$ does not
satisfy the separation axiom on the class of tree-sibling model
phylogenetic networks recalled above.  See \cite{cardona.ea:07a} for a
detailed discussion of this issue.

Two other dissimilarity measures considered in \cite{nakhleh:phd04,%
nakhleh.ea:TR0406,nakhleh.ea:03psb} are based on the representation of
a rooted DAG by means of its \emph{induced subtrees}: the phylogenetic
trees with the same root and the same leaves as the network that are
obtained by taking a spanning subtree of the network and then
contracting elementary paths into nodes.  For every rooted DAG $N$, let $\mathcal{T}(N)$
denote the set of all its induced subtrees, and $\mathcal{C}(N)$ the
set of all clusters of nodes of these induced subtrees.

Then, for every two rooted DAGs $N_{1}=(V_{1},E_{1})$ and
$N_{2}=(V_{2},E_{2})$ labeled in the same set $S$, the authors define:
\begin{itemize}
\item $m^{tree}(N_{1},N_{2})$ as the weight of a minimum weight edge
cover of the complete bipartite graph with nodes
$\mathcal{T}(N_{1})\sqcup \mathcal{T}(N_{2})$ and edge weights the
value of the  Robinson-Foulds metric  between the pairs of induced
subtrees of $N_1$ and $N_2 $ connected by each edge.

\item $m^{sp}(N_{1},N_{2})$   as $m_{\Upsilon}$, replacing
$\Upsilon$ by $\mathcal{C}$:
$$
m^{sp}(N_{1},N_{2})=\frac{1}{2}\Bigl(\frac{| \mathcal{C}(N_{1})\setminus
\mathcal{C}(N_{2})|}{|E_{1}|}+\frac{| \mathcal{C}(N_{2})\setminus
\mathcal{C}(N_{1})|}{|E_{2}|}\Bigr).
$$
\end{itemize}
These measures do not satisfy the separation axiom on the class of
tree-sibling phylogenetic networks: see, for instance,
\cite[Fig.~6.8]{nakhleh:phd04}.  On the positive side, Nakhleh et al
prove in \cite[\S 6.4]{nakhleh:phd04} and \cite[\S
5]{nakhleh.ea:TR0406} that they are distances on the subclass of
time-consistent binary galled trees.  But it can be easily checked that on
arbitrary galled trees they do not define distances either: see, for
instance, Fig.~\ref{fig:treemetric}.

\section{Tree-child phylogenetic networks}
\label{sec:TC}

Since in this paper we are not interested in the reconstruction of
networks, for the sake of generality we assume the most general notion
of \emph{phylogenetic network} on a set $S$ of taxa: any rooted
$S$-DAG. So, its hybrid nodes can have any in-degree greater than one
and any out-degree, and its tree-nodes can have any out-degree.  In
particular, they may contain hybrid leaves and out-degree 1 tree
nodes.

We shall introduce two comparison methods on a
specific subclass of such networks.

\begin{definition}
A phylogenetic network satisfies the \emph{tree-child condition},
or it is a \emph{tree-child phylogenetic network}, when every internal node
has at least one tree child.  
\end{definition}

Tree-child phylogenetic networks can be understood thus as general
models of reticulated evolution where every species other that the
extant ones, represented by the leaves, has some descendant through
mutation.  This slightly strengthens the condition imposed on
phylogenetic networks in \cite{nakhleh.ea:TR0209}, where tree nodes
had to have at least one tree child, because we also require internal
hybrid nodes to have some tree child.  So, if hybrid nodes are further
imposed to have exactly one child (as for instance in the definition
of model phylogenetic network recalled in \S \ref{subsec:phnet}), this
node must be a tree node: this corresponds to the interpretation of
hybrid nodes not as individuals but as recombination events, producing
a new individual represented by their only child.  On the other hand,
if hybrid nodes represent individuals, then a hybrid node with all its
children hybrid corresponds to a hybrid individual that hybridizes
before undergoing a speciation event, a scenario, according to
\cite{nakhleh.ea:TR0209}, that ``almost never arises in reality.''

The following result gives two other alternative characterizations of
tree-child phylogenetic networks in terms of their strict and tree
clusters.

\begin{lemma}\label{lem:TC}
The following three conditions are equivalent for every phylogenetic
network $N=(V,E)$:
\begin{enumerate}[(a)]
\item $N$ is tree-child.  
\item $T_{L}(v)\neq \emptyset$ for every node $v\in V\setminus V_L$.  
\item $A_{L}(v)\neq
\emptyset$ for every node $v\in V$.
\end{enumerate}
\end{lemma}

\begin{proof}
(a)$\Longrightarrow$(b): Given any node $v$  other than a leaf, we can
construct a tree path by successively taking tree children.  This path
must necessarily end in a leaf that, by definition, belongs to
$T_{L}(v)$.

(b)$\Longrightarrow$(c): If $v\notin V_{L}$, then, by Corollary
\ref{cor:unicity-path}, $\emptyset\neq T_{L}(v)\subseteq A_{L}(v)$,
while if $v\in V_{L}$, then, by definition, $v\in A_{L}(v)$.

(c)$\Longrightarrow$(a): Let $v$ be any internal node.  We want to
prove that if $A_L(v)\neq \emptyset$, then $v$ has a tree child.  So,
let $s\in A_L(v)$, and consider the set $W$ of children of $v$ that
are ancestors of $s$: it is non-empty, because $s$ must be a
descendant of some child of $v$.  Let $w$ be a maximal element of $W$
with respect to the path ordering on $N$.  If $w$ is a tree node, we
are done.  Otherwise, let $v'$ be a parent of $w$ different from $v$.
Let $r\pathgr v'$ be any path from a root $r$ to $v'$.  Concatenating
this path with the arc $(v',w)$ and any path $w\pathgr s$, we get a
path $r\pathgr s$.  Since $s\in A_L(v)$, this path must contain $v$,
and then, since $N$ is acyclic, $v$ must be contained in the path $r
\pathgr v'$.  Let $w'$ be the node that follows $v$ in this path.
This node $w'$ is a child of $v$ and there exists a non-trivial path
$w'\pathgr w$ (through $v'$), which makes $w'$ also an ancestor of
$s$.  But then $w'\in W$ and $w'>w$, which contradicts the maximality
assumption on $w$.  \qed
\end{proof}

Next lemma shows that tree-child phylogenetic networks are a more
general model of evolution under mutation and recombination than the
galled trees.

\begin{lemma}
\label{lem:galled}
Every rooted galled tree is a tree-child phylogenetic network.
\end{lemma}

\begin{figure}[htb]
\begin{center}
\begin{tikzpicture}[thick,>=stealth,xscale=0.7,yscale=0.8]
\draw(0,0) node[tre] (ab) {}; \draw (ab.north) node[above] {$lca(a,b)$};
\draw(2,0) node[tre] (bup) {}; \draw (bup.north) node[above] {$lca(b,u')$};
\draw(-1,-1) node[tre] (a) {}; \etq a
\draw(1,-1) node[hyb] (b) {}; \etq b
\draw(0,-2) node[hyb] (u) {}; \etq u
\draw(3,-2) node[tre] (u') {}; \etq {u'}
\draw(1.5,-3) node[hyb] (v_1) {}; \etq {v_1}
\draw[->,dashed](ab)--(a);
\draw[->,dashed](ab)--(b);
\draw[->,dashed](bup)--(b);
\draw[->,dashed](bup)--(u');
\draw[->](a)--(u);
\draw[->](b)--(u);
\draw[->](u)--(v_1);
\draw[->](u')--(v_1);
\end{tikzpicture}
\qquad
\begin{tikzpicture}[thick,>=stealth]
\draw(0,0) node[tre] (wu1) {};\draw (wu1.north) node[above] {$lca(w,u_1)$};
\draw(2,0) node[tre] (wu2) {};\draw (wu2.north) node[above] {$lca(w,u_2)$};
\draw(1,-1) node[hyb] (w) {}; \etq w
\draw(-1,-2) node[tre] (u_1) {}; \etq{u_1}
\draw(3,-2) node[tre] (u_2) {}; \etq{u_2}
\draw(1,-2) node[tre] (u) {}; \etq{u}
\draw(0,-3) node[hyb] (v_1) {};\etq {v_1}
\draw(2,-3) node[hyb] (v_2) {};\etq {v_2}
\draw[->,dashed] (wu1)--(w);
\draw[->,dashed] (wu2)--(w);
\draw[->,dashed] (wu1)--(u_1);
\draw[->,dashed] (wu2)--(u_2);
\draw[->](w)--(u);
\draw[->](u_1)--(v_1);
\draw[->](u)--(v_1);
\draw[->](u_2)--(v_2);
\draw[->](u)--(v_2);
\end{tikzpicture}
\end{center}
\caption{\label{fig:galled}
(a) In a galled tree, a hybrid node cannot have a hybrid child.  (b)
In a galled tree, a non-root tree node  cannot have two hybrid children.}
\end{figure}
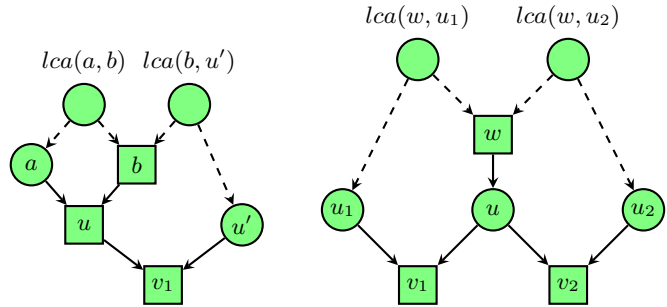

\begin{proof}
Let $N=(V,E)$ be a galled tree.  If $N$ does not satisfy the
tree-child condition, then it contains an internal node $u \in V$ with
all its children $v_{1},\ldots,v_{k} \in V$ hybrid.

The node $u$ cannot be hybrid, because in galled trees a hybrid node
cannot have any hybrid children.  Indeed, assume that $u$ has two
parents $a,b$, and let $u'$ be the other parent of the child $v_{1}$
of $v$.  Let $x$ be the least common ancestor of $a$ and $b$, and $y$
the least common ancestor of $b$ and $u'$.  Then the recombination
cycles defined by the paths $(x,\ldots,a,u)$ and $(x,\ldots,b,u)$, on
the one hand, and $(y,\ldots,b,u,v_{1})$ and $(y,\ldots,u',v_{1})$ on
the other hand, share the arc $(b,u)$, contradicting the hypothesis
that $N$ is a galled tree.  See Fig.~\ref{fig:galled}.(a).

Thus, $u$ is a tree node. In this case, 
$k\geq 2$, because  galled trees cannot have out-degree 1 tree nodes.
Now, if $u$ is the root of $N$, then $A_{L}(u)=V_{L}\neq \emptyset$
and hence, by the proof of the implication (c)$\Rightarrow$(a) in 
Lemma \ref{lem:TC}, it has some tree child.
If, on the contrary, $u$ is not the root of $N$, let $w$ be its parent and $u_{1}$
and $u_{2}$  the parents other than $u$ of $v_{1}$ and $v_{2}$,
respectively. Let $x_{1}$ be the least common ancestor of $w$ and
$u_{1}$, and $x_{2}$ the least common ancestor of $w$ and $u_{2}$.
Then the recombination cycles defined by the paths
$(x_{1},\ldots,u_{1},v_{1})$ and $(x_{1},\ldots,w,u,v_{1})$, on the
one hand, and $(x_{2},\ldots,u_{2},v_{2})$ and
$(x_{2},\ldots,w,u,v_{2})$, on the other hand, share the arc $(w,u)$,
contradicting again the hypothesis that $N$ is a galled tree.  See
Fig.~\ref{fig:galled}.(b).  \qed
\end{proof}

\begin{remark}
Not every tree-child phylogenetic network is a
galled tree: see, for instance, the tree-child phylogenetic network in 
Fig.~\ref{fig:exemple1}.
\end{remark}

We provide now some upper bounds on the number of nodes in a
tree-child phylogenetic network.

\begin{proposition}\label{lem:bound}
 Let $N=(V,E)$ be a tree-child phylogenetic network with $n$ leaves.
 \begin{enumerate}[(a)] 
\item $|V_{H}|\leq n-1$.

\item If $N$  has no out-degree 1 tree node, then $\displaystyle
|V|\leq 2n-1+\sum_{v\in V_{H}} d_{i}(v)$.

\item If $N$  has no out-degree 1 tree node and if
$m=\max\{d_{i}(v)\mid v\in V_{H}\}$, then $|V|\leq (m+2)(n-1)+1$.
 \end{enumerate}
\end{proposition}

\begin{proof}
(a) Let $r$ be the root of $N$.  Consider a mapping $t:V\smallsetminus
V_{L}\to V_T\smallsetminus\{r\}$ that assigns to every internal node
one of its tree children; since tree nodes have a single parent, this
mapping is injective.  Then, $|V|-|V_{L}|\leq |V_T|-1$ and, since
$|V|=|V_{H}|+|V_T|$ and $|V_{L}|=n$, 
$$
|V_{H}|=|V|-|V_{T}|\leq |V_{L}|-1=n-1.
$$

(b) For every $j\geq 2$, let $V_{H,j}$ be the set of hybrid nodes with
in-degree $j$.  If, for every hybrid node $v$, we remove from $N$ a
set of $d_{i}(v)-1$ arcs with head $v$, we obtain a tree with set of
nodes $V$ and set of leaves $V_{L}$ (no internal node of $N$ becomes a
leaf, because when we remove an arc $e$, since it is a hybridization arc,
there still remains some tree arc with the same tail as $e$).  Now, in
this tree there will be at most $\sum_{v\in V_{H}} d_{i}(v)$ nodes
with in and out-degree 1: $N$ did not have any out-degree 1 tree node
and, in the worst case, when we remove the $d_{i}(v)-1$ arcs with head
$v$, this node and the tails of the removed arcs become nodes of
in and out-degree 1.  Since, in a tree, the number of nodes is smaller
than twice the number of leaves plus the number of nodes with in and
out-degree 1, the inequality in the statement follows.

(c) If $m=\max\{d_{i}(v)\mid v\in V_{H}\}$, then $\sum_{v\in V_{H}}
d_{i}(v)\leq m|V_{H}|$. Then, combining (a) and (b),
$$
|V|\leq 2n-1+\sum_{v\in V_{H}} d_{i}(v)\leq
2n-1+m|V_{H}|\leq 2n-1+m(n-1)=(m+2)(n-1)+1,
$$
as we claimed.
\qed
\end{proof}

The upper bounds in Lemma \ref{lem:bound} are sharp, as there exist
tree-child phylogenetic networks for which these inequalities are
equalities: for $n=1$, point (a) entails that $N$ is a tree, and (b) and (c) then simply say that $N$ consists only of one node; for $n\geq 2$,
 see the next example.  In particular, for every number $n\geq 2$ of leaves, there exist
arbitrarily large tree-child phylogenetic networks without out-degree 1 tree nodes with $n$ leaves.  Of course, if we do not forbid
out-degree 1 tree nodes, then there exists no upper bound on the
number of nodes of the network.

\begin{figure}[htb]
\begin{center}
    \begin{tikzpicture}[thick,>=stealth,xscale=0.45,yscale=0.5]
   \draw(0,0) node[tre] (v_{1,3}) {};  \etq{v_{1,3}}
    \draw(-1.333,-4) node[trenou] (v_{1,2}) {}; \etq{v_{1,2}}
    \draw(-2.666,-8) node[trenou] (v_{1,1}) {}; \etq{v_{1,1}}
    \draw(-4,-12) node[tre] (1) {}; \etq{1}
    \draw(3,-2) node[hybnou] (h_{2}) {}; \etq{h_{2}}
    \draw(6,-4) node[tre] (v_{2,3}) {}; \etq{v_{2,3}}
    \draw(5,-6.666) node[trenou] (v_{2,2}) {}; \etq{v_{2,2}}
    \draw(4,-9.333) node[trenou] (v_{2,1}) {}; \etq{v_{2,1}}
    \draw(3,-12) node[tre] (2) {}; \etq{2}
    \draw(9,-6) node[hybnou] (h_{3}) {}; \etq{h_{3}}
    \draw(12,-8) node[tre] (v_{3,3}) {}; \etq{v_{3,3}}
      \draw(11,-9.333) node[trenou] (v_{3,2}) {}; \etq{v_{3,2}}
      \draw(10,-10.666) node[trenou] (v_{3,1}) {}; \etq{v_{3,1}}
      \draw(9,-12) node[tre] (3) {}; \etq{3}
      \draw(15,-10) node[hybnou] (h_{4}) {}; \etq{h_{4}}
      \draw(18,-12) node[tre] (4) {}; \etq{4}
    \draw[->] (v_{1,3})--(v_{1,2});
    \draw[->] (v_{1,2})--(v_{1,1});
    \draw[->] (v_{1,1})--(1);
    \draw[->] (v_{1,3})--(h_{2});
    \draw[->] (h_{2})--(v_{2,3});
    \draw[->] (v_{2,3})--(v_{2,2});
    \draw[->] (v_{2,2})--(v_{2,1});
    \draw[->] (v_{2,1})--(2);
    \draw[->] (v_{2,3})--(h_{3});
    \draw[->] (h_{3})--(v_{3,3});
    \draw[->] (v_{3,3})--(v_{3,2});
    \draw[->] (v_{3,2})--(v_{3,1});
    \draw[->] (v_{3,1})--(3);
    \draw[->] (v_{3,3})--(h_{4});
    \draw[->] (h_{4})--(4);
    \draw[->] (v_{1,3})--(h_{2});
    \draw[->] (v_{1,2})--(h_{2});
    \draw[->] (v_{1,1})--(h_{2});
    \draw[->] (v_{2,3})--(h_{3});
    \draw[->] (v_{2,2})--(h_{3});
    \draw[->] (v_{2,1})--(h_{3});
    \draw[->] (v_{3,3})--(h_{4});
    \draw[->] (v_{3,2})--(h_{4});
    \draw[->] (v_{3,1})--(h_{4});
  \end{tikzpicture}
\end{center}
\caption{\label{fig:bound}
A tree-child phylogenetic network with 4 leaves and $5\cdot 3+1$ nodes.}
\end{figure}
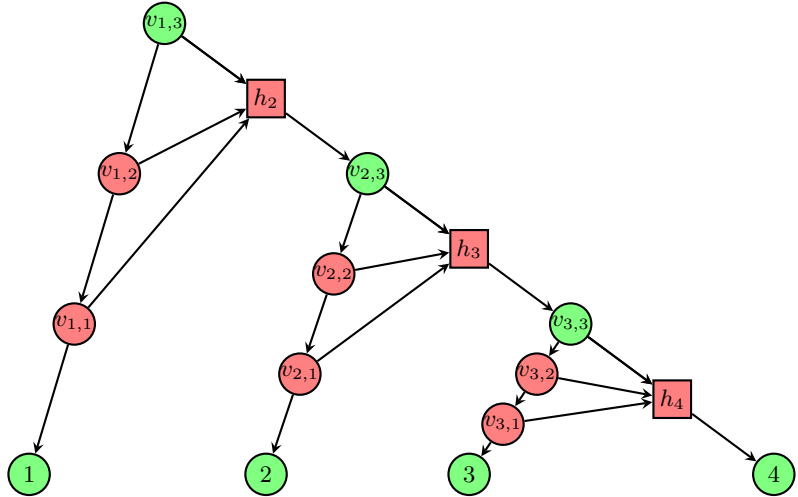

\begin{example}\label{ex:bignetwork}
Let $T$ be the `comb-like' binary phylogenetic tree labeled in
$\{1,\ldots,n\}$ described by the Newick string
$$
(1,(2,(3,\ldots,(n-1,n)\ldots ))),
$$
and let us fix a positive integer number $m\geq 2$.

For every $i=1,\ldots,n-1$ let us call $v_{i,m}$ the parent of the leaf
$i$: to simplify the language, set $v_{n,m}=n$.  Notice that $v_{1,m}$
is the root of the tree.  Now, for every $i=1,\ldots,n-1$, split the
arc $(v_{i,m},i)$ into a path of length $m$,
$$
(v_{i,m},v_{i,m-1},\ldots,v_{i,1},i),
$$
split the arc $(v_{i,m},v_{i+1,m})$ into a  path of length 2,
$$
(v_{i,m},h_{i+1},v_{i+1,m}),
$$
and, for every $i=1,\ldots,n-1$ and $j=1,\ldots,m-1$, add an arc
$(v_{i,j},h_{i+1})$. Fig.~\ref{fig:bound} displays\footnote{Henceforth, in
graphical representations of phylogenetic networks, and of DAGs in
general, hybrid nodes are represented by squares and tree nodes by
circles.} this construction
for $n=4$ and $m=3$. 

The original binary tree had $2n-1$ nodes, and we have added
$(m-1)(n-1)$ new tree nodes and $n-1$ hybrid nodes (of in-degree $m$).
Therefore, the resulting tree-child phylogenetic network has
$(m+2)(n-1)+1$ nodes.
\end{example}

\begin{figure}[htb]
\begin{center}
            \begin{tikzpicture}[thick,>=stealth,scale=1.7]
              \draw(0,0) node[tre] (r) {}; \etq r
              \draw(0.5,-1) node[tre] (u) {}; \etq u
              \draw(-1,-2) node[tre] (1) {}; \etq 1
              \draw(0,-2) node[tre] (2) {}; \etq 2
              \draw(1,-2) node[tre] (3) {}; \etq 3      
            \draw[->] (r)--(u);
            \draw[->] (r)--(1);
            \draw[->] (u)--(3);
            \draw[->] (u)--(2);
            \end{tikzpicture}
  \qquad\qquad
  \begin{tikzpicture}[thick,>=stealth,scale=0.85]
      \draw(0,2) node[tre] (r) {}; \etq r
        \draw(0,1) node[tre] (b) {}; \etq b
        \draw(1,0) node[tre] (a) {}; \etq a
        \draw(-1,-1) node[hyb] (A) {}; \etq A
         \draw(-1,-2) node[tre] (2) {}; \etq 2
        \draw(1,-2) node[tre] (3) {}; \etq 3
        \draw(-3,-2) node[tre] (1) {}; \etq 1
        \draw[->] (r)--(b);
        \draw[->] (r)--(1);
        \draw[->] (b)--(a);
        \draw[->] (b)--(A);
        \draw[->] (a)--(A);
        \draw[->] (A)--(2);
        \draw[->] (a)--(3);    
  \end{tikzpicture}
\end{center}
\caption{\label{fig:treemetric} 
A tree and a galled tree with the same sets of induced subtrees.}
\end{figure}
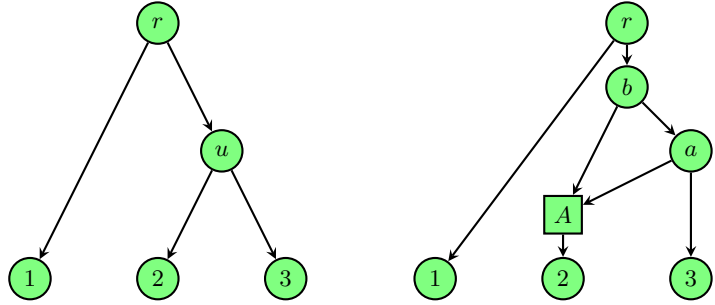

In the next sections, we define a distance on the class of all
tree-child phylogenetic networks.  It is convenient thus to remember
here that the tripartition metrics $m_\theta$ or $m_\Psi$ recalled in
\S \ref{subsec:metr} do not define a  distance on this class,
because there exist pairs of non-isomorphic tree-child phylogenetic
networks on the same set of taxa with the same sets of enriched tripartitions: for instance, 
the networks depicted in  Figs.~\ref{fig:exemple1} and~\ref{fig:exemple1bis} below (see
 \cite{cardona.ea:07a} for details).  As far as the
metrics $m^{tree}$ and $m^{sp}$ also recalled in \S \ref{subsec:metr}
goes, they do not define either distances on the class of all
tree-child phylogenetic networks, because there also exist pairs of
non-isomorphic tree-child phylogenetic networks on the same set of
taxa with the same sets of induced subtrees.  For instance, the tree
and the galled tree depicted in Fig.~\ref{fig:treemetric} have the
same sets of induced subtrees, namely the tree itself, and hence the
same sets of clusters of induced subtrees.

\section{The $\mu$-representation of tree-child phylogenetic networks}
\label{sec:mu-representation}

Let us fix henceforth a set of labels $\LL=\{l_1,\dots,l_n\}$: unless
otherwise stated, all DAGs appearing henceforth are assumed to be
labeled in $\LL$, usually without any further notice.

Let $N=(V,E)$ be an $\LL$-DAG. For every node $u\in V$ and for every
$i=1,\ldots,n$, we denote by $m_i(u)$ the number of different paths
from $u$ to the leaf $l_i$.  We define the \emph{path-multiplicity
vector}, or simply \emph{$\mu$-vector} for short, of $u\in V$ as
$$
\mu(u)=(m_1(u),\dots,m_n(u));
$$
that is, $\mu(u)$ is the $n$-tuple holding the number of paths from
$u$ to each leaf of the graph.

To simplify the notations, we shall denote henceforth by $\delta_i^{(n)}$ the unit
 vector 
 $$
 (\underbrace{0,\ldots,0,\stackrel{i}{1_{{}}},0,\ldots,0}_n).
 $$

\begin{lemma}\label{lem:comp-mu}
Let $u\in V$ be any node of an $\LL$-DAG $N=(V,E)$.
\begin{enumerate}[(a)]
\item If $u=l_i\in V_{L}$, then
$\mu(u)=\delta_i^{(n)}$.

\item If $u\notin V_{L}$ and $\child(u)=\{v_1,\dots,v_k\}$, then
$\mu(u)=\mu(v_1)+\dots+\mu(v_k)$.
\end{enumerate}
\end{lemma}

\begin{proof}
The statement for leaves is trivial.  When $u\in V\setminus V_{L}$, by deleting
or prepending $u$ we get, for every $i=1,\ldots,n$, a bijection
$$
\{\mbox{paths }u\pathgr l_{i}\} \leftrightarrow \bigsqcup_{1\leq j\leq
k}\{\mbox{paths }v_{j}\pathgr l_{i}\}
$$
which clearly implies the statement in this case.\qed
\end{proof}

\begin{remark}
\label{rem:1}
If $v\in \child(u)$, then $\mu(u)=\mu(v)$ if
and only if $v$ is the only child of $u$: any other child would
contribute something to $\mu(u)$.
\end{remark}

Lemma~\ref{lem:comp-mu} implies the simple Algorithm~\ref{alg0} to compute the $\mu$-vectors of the nodes of an $S$-DAG in polynomial time.
Since the height of the nodes can be computed in $O(n+|E|)$ time, it takes $O(n|E|)$ time to compute $\mu(N)$ on an $S$-DAG $N=(V,E)$ with $n$ leaves.

\begin{algorithm}
\label{alg0}
Given an $S$-DAG $N=(V,E)$, compute $\mu(N)$.
\upshape
\begin{tabbing}
\quad \=\quad \=\quad \=\quad \=\quad \=\quad \=\quad \kill
\textbf{begin}\\
\> \textbf{for} $i=1,\ldots,n$ \textbf{do} \\
\> \> set $\mu(l_{i})=\delta_i^{(n)}$ \\
\> sort $V\setminus V_{L}$ increasingly on \emph{height} \\
\> \textbf{for each} $x \in V\setminus V_{L}$ \textbf{do} \\
\> \> let $y_{1},\ldots,y_{k} \in V$ be the children of $x$ \\
\> \> set $\mu(x)=\mu(y_{1})+\cdots +\mu(y_{k})$\\
\textbf{end}
\end{tabbing}
\end{algorithm}

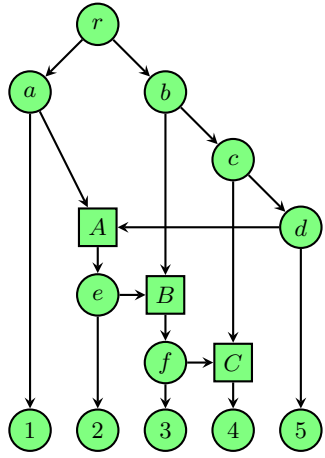
\begin{figure}[htb]
\begin{center}
    \begin{tikzpicture}[thick,>=stealth,scale=0.9]
\draw (3,7) node[tre] (r) {}; \etq r
\draw (2,6) node[tre] (a) {}; \etq a
\draw (4,6) node[tre] (b) {}; \etq b
\draw (2,1) node[tre] (1) {}; \etq 1
\draw (5,5) node[tre] (c) {}; \etq c
\draw (3,4) node[hyb] (A) {}; \etq A
\draw (6,4) node[tre] (d) {}; \etq d
\draw (3,3) node[tre] (e) {}; \etq e
\draw (4,3) node[hyb] (B) {}; \etq B
\draw (6,1) node[tre] (5) {}; \etq 5
\draw (3,1) node[tre] (2) {}; \etq 2
\draw (4,2) node[tre] (f) {}; \etq f
\draw (5,2) node[hyb] (C) {}; \etq C
\draw (4,1) node[tre] (3) {}; \etq 3
\draw (5,1) node[tre] (4) {}; \etq 4
\draw [->](r)--(a);
\draw [->](r)--(b);
\draw [->](a)--(1);
\draw [->](a)--(A);
\draw [->](b)--(B);
\draw [->](b)--(c);
\draw [->](c)--(C);
\draw [->](c)--(d);
\draw [->](A)--(e);
\draw [->](d)--(A);
\draw [->](d)--(5);
\draw [->](e)--(2);
\draw [->](e)--(B);
\draw [->](B)--(f);
\draw [->](f)--(3);
\draw [->](f)--(C);
\draw [->](C)--(4);
\end{tikzpicture}
\end{center}
\caption{\label{fig:exemple1}
The tree-child phylogenetic network used in Example~\ref{ex:mu-repr1}.}
\end{figure}

\begin{example}
\label{ex:mu-repr1}
Consider the tree-child phylogenetic network depicted in
Fig.~\ref{fig:exemple1}.  Table~\ref{tab:taula1} gives the $\mu$-vectors
of its nodes, sorted increasingly by their heights.
\end{example}

\begin{table}[htb]
   \centering
\begin{tabular}{|c|c|c||c|c|c||c|c|c|}
\hline
node & height & $\mu$-vector  & node  & height & $\mu$-vector & node  & height & $\mu$-vector  \\
\hline
$1$ & 0 &{}\   $(1,0,0,0,0)$\ {}  &
$C$ & 1 &{}\   $(0,0,0,1,0)$\ {}  &
$a$ & 6 &{}\  $(1,1,1,1,0)$\ {} \\
\hline
$2$ & 0 & $(0,1,0,0,0)$ &
$f$ & 2 & {}\ $(0,0,1,1,0)$\ {}  &
$d$ & 6 & $(0,1,1,1,1)$\\
\hline
$3$ & 0 & $(0,0,1,0,0)$ &
$B$ & 3 & $(0,0,1,1,0)$ &
$c$ & 7 & $(0,1,1,2,1)$\\
\hline
$4$ & 0 & $(0,0,0,1,0)$ &
$e$ & 4 & $(0,1,1,1,0)$ &
$b$ & 8 & $(0,1,2,3,1)$\\
\hline
$5$ & 0 & $(0,0,0,0,1)$ &
 $A$ & 5 & $(0,1,1,1,0)$ &
$r$ & 9 & $(1,2,3,4,1)$\\
\hline
 \end{tabular} 
 \smallskip
\caption{$\mu$-vectors of the nodes of the network depicted in
Fig.~\ref{fig:exemple1}.}
   \label{tab:taula1}
\end{table}

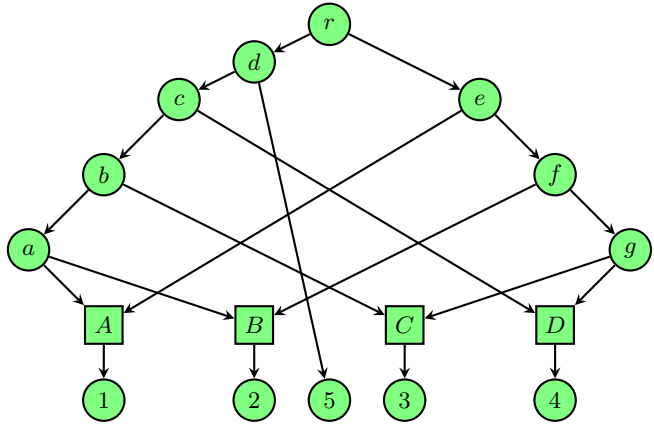
\begin{figure}[htb]
\begin{center}
    \begin{tikzpicture}[thick,>=stealth]
    \draw(0,0) node[tre] (r) {}; \etq r
    \draw(-2,-1) node[tre] (c) {};\etq c
    \draw(-3,-2) node[tre] (b) {};\etq b
    \draw(-4,-3) node[tre] (a) {};\etq a
    \draw(2,-1) node[tre] (e) {};\etq e
    \draw(3,-2) node[tre] (f) {};\etq f
    \draw(4,-3) node[tre] (g) {};\etq g
    \draw(-3,-4) node[hyb] (A) {};\etq A
    \draw(-1,-4) node[hyb] (B) {};\etq B
    \draw(1,-4) node[hyb] (C) {};\etq C
    \draw(3,-4) node[hyb] (D) {};\etq D
    \draw(-3,-5) node[tre] (1) {};\etq 1
    \draw(-1,-5) node[tre] (2) {};\etq 2
    \draw(1,-5) node[tre] (3) {};\etq 3
    \draw(3,-5) node[tre] (4) {};\etq 4
    \draw(-1,-0.5) node[tre] (d) {};\etq d
    \draw(0,-5) node[tre] (5) {};\etq 5
    \draw[->](r)--(d);
    \draw[->](r)--(e);
    \draw[->](d)--(c);
    \draw[->](d)--(5);
    \draw[->](c)--(b);
    \draw[->](c)--(D);
    \draw[->](e)--(A);
    \draw[->](e)--(f);
    \draw[->](b)--(a);
    \draw[->](b)--(C);
    \draw[->](f)--(g);
    \draw[->](f)--(B);
    \draw[->](a)--(A);
    \draw[->](a)--(B);
    \draw[->](g)--(C);
    \draw[->](g)--(D);
    \draw[->](A)--(1);
    \draw[->](B)--(2);
    \draw[->](C)--(3);
    \draw[->](D)--(4);
  \end{tikzpicture}
\end{center}
\caption{\label{fig:contr1}The tree-sibling, non-tree-child,
phylogenetic network used in Example~\ref{ex:mu-repr2}.}
\end{figure}

\begin{example}
\label{ex:mu-repr2}
Consider the phylogenetic network depicted in Fig.~\ref{fig:contr1}.
Table~\ref{tab:taula2} gives the $\mu$-vectors of its nodes sorted
increasingly by their heights.
\end{example}

\begin{table}[htb]
   \centering
\begin{tabular}{|c|c|c||c|c|c||c|c|c|}
\hline
node & height & $\mu$-vector  & node  & height & $\mu$-vector & node  & height & $\mu$-vector  \\
\hline
$1$ & 0 &{}\   $(1,0,0,0,0)$\ {}  &
$B$ & 1 & {}\   $(0,1,0,0,0)$\ {}   &
$f$ & 3 & {}\   $(0,1,1,1,0)$\ {}  
\\
\hline
$2$ & 0 & $(0,1,0,0,0)$ &
$C$ & 1 & $(0,0,1,0,0)$ &
$c$ & 4 & $(1,1,1,1,0)$ 
\\
\hline
$3$ & 0 & $(0,0,1,0,0)$ &
$D$ & 1 & $(0,0,0,1,0)$ &
$e$ & 4 & $(1,1,1,1,0)$ 
\\
\hline
$4$ & 0 & $(0,0,0,1,0)$ &
$a$ & 2 & $(1,1,0,0,0)$ &
$d$ & 5 & $(1,1,1,1,1)$ 
\\
\hline
$5$ & 0 & $(0,0,0,0,1)$ &
$g$ & 2 &$(0,0,1,1,0)$ &
$r$ &  6 & $(2,2,2,2,1)$ 
\\
\hline
$A$ & 1 & {}\ $(1,0,0,0,0)$\ {} &
$b$ & 3 &  {}\ $(1,1,1,0,0)$\ {} 
\\
\cline{1-6}
 \end{tabular} 
 \smallskip
\caption{$\mu$-vectors of the nodes of the network depicted in
Fig.~\ref{fig:contr1}.}
   \label{tab:taula2}
\end{table}

\begin{definition}
The \emph{$\mu$-representation} of a DAG $N=(V,E)$ is the multiset
$\mu(N)$ of $\mu$-vectors of its nodes: its elements are the vectors
$\mu(u)$ with $u\in V$, and each one appears in $\mu(N)$ as many times
as the number of nodes having it as $\mu$-vector.
\end{definition}

It turns out that a tree-child phylogenetic network can be singled out
up to isomorphism among all tree-child phylogenetic network by means
of its $\mu$-representation (Thm.~\ref{thm-isomorphism}).  Before
proceeding with the proof of this fact, we establish several auxiliary
results.

The following lemma shows that the path ordering on a tree-child
phylogenetic network is almost determined by its $\mu$-representation.
In it, and henceforth, the order $\geq$ considered between $\mu$-vectors is the
\emph{product partial order} on $\NN^n$:
$$
(m_1,\dots,m_n)\ge (m'_1,\dots,m'_n)\quad \iff\quad m_i\ge m'_i \mbox{ for
every $i=1,\dots,n$.}
$$

\begin{lemma}\label{lemanou}
Let $N=(V,E)$ be a tree-child phylogenetic network.

\begin{enumerate}[(a)]
\item If there exists a path $u\pathgr v$, then $\mu(u)\ge \mu(v)$.

\item If $\mu(u)>\mu(v)$, then there exists a path $u\pathgr v$.

\item If $\mu(u)=\mu(v)$, then $u$ and $v$ are connected by an
elementary path.

\end{enumerate}
\end{lemma}

\begin{proof}
Assertion (a) is a straightforward consequence of Lemma
\ref{lem:comp-mu}.

As far as (b) and (c) goes, let us assume for the moment that
$\mu(u)\geq \mu(v)$ and let $l_i\in T_{L}(v)$; in particular,
$m_{i}(v)\geq 1$.  Then, $m_i(u)\geq m_i(v)$, and therefore there
exists also a path $u\pathgr l_i$.  Now, consider the tree path
$v\pathgr l_i$.  By Lemma \ref{lem:unicity-path}, it must happen that
either the path $u\pathgr l_i$ contains the path $v\pathgr l_i$ or
vice versa, and therefore $u$ and $v$ are connected by a path.

If $\mu(u)>\mu(v)$, by (a) there cannot exist any path $v\pathgr u$, and
therefore there exists a path $u\pathgr v$: this proves (b).

On the other hand, if $\mu(u)=\mu(v)$, nothing prevents the existence
of a path $u\pathgr v$ or a path $v\pathgr u$.  To fix ideas, assume that
there exists a path $u\pathgr v$, say $(u,v_{1},\ldots,v_{k-1},v)$.
Since, by (a),
$$
\mu(u)\geq \mu(v_{1})\geq \cdots \geq \mu(v_{k-1})\geq \mu(v)=\mu(u),
$$
we conclude that
$$
\mu(u)= \mu(v_{1})= \cdots = \mu(v_{k-1})= \mu(v).
$$
As we noticed in Remark \ref{rem:1}, this implies that each one of
$u,v_{1},\ldots,v_{k-1}$ has only one child, the node that follows it
in this path.  Therefore the path $u\pathgr v$ is elementary.
The same argument shows that if it is the path $v\pathgr u$ which
exists, then it is elementary.  \qed
\end{proof}

\begin{remark}
Assertion (a) in the previous lemma holds for every DAG, but
assertions (b) and (c) need not be true if the DAG $N$ does not
satisfy the tree-child condition, even if $N$ is a binary tree-sibling
phylogenetic network.  Indeed, consider the phylogenetic network
described in Fig.~\ref{fig:contr1}.  We have in it that
$$
\begin{array}{ll}
\mu(c)& =(1,1,1,1,0) = \mu(e) \\
\mu(d)& =(1,1,1,1,1) 
\end{array}
$$
Then, $c$ and $d$ have the same $\mu$-vectors but they are not
connected by any path, yielding a counterexample to (c).  And
$\mu(d)>\mu(e)$, but there is no path $d\pathgr
e$, which yields a counterexample to (b).
\end{remark}

%
%
%
%

The next two lemmas show how to recover the children of a node in a tree-child
phylogenetic network  from the knowledge of the
$\mu$-representation of the network. 

\begin{lemma}\label{lem:Mu}
Let $N$ be any DAG. Let $u\in V$ be any internal node, and
let
$$
M_{u}=\{w\in V\mid u>w\}.
$$
Then, $M_{u}$ has maximal elements, and all of them are children of
$u$.
\end{lemma}

\begin{proof}
The set $M_{u}$ is non-empty, since $u$ is not a leaf and every
descendant of $u$ is in $M_{u}$.  Since $M_{u}$ is finite, it has
maximal elements.  Let $v$ be any such a maximal element.  Since
$u>v$, there exists a non trivial path $u\pathgr v$.  If this path
passes through some other node $w$, then $u>w>v$, against the
assumption that $v$ is maximal in $M_{u}$.  Therefore, the path
$u\pathgr v$ has length 1 and $v$ is a child of $u$.  \qed
\end{proof}

The maximal elements of $M_{u}$ are exactly the children of $u$ such
that the arc $(u,v)$ is the only path $u\pathgr v$.  This includes all
tree children of $u$, as well as all hybrid children $v$ of $u$ such
that no other parent of $v$ is a descendant of $u$.  In time consistent
phylogenetic networks, this covers all children of $u$.  But in
arbitrary tree-child phylogenetic networks, this need not cover all of
them.  Consider for instance the right-hand side tree-child phylogenetic network
in  Figure \ref{fig:treemetric}.  In it, the only maximal element
of $M_{b}$ is $a$, but $A$ is also a child of $b$.

%

\begin{lemma}\label{Prop:Muv1vk}
Let $N$ be a tree-child phylogenetic network.  Let $u\in V$ be any
internal node and $v_1,\dots,v_k$ some of its children.
\begin{enumerate}[(a)]
\item If $\mu(u)=\mu(v_1)+\dots+\mu(v_k)$, then $u$ has no other
children.

\item If $\mu(u)>\mu(v_1)+\dots+\mu(v_k)$, let
$$
M_{u,v_{1},\ldots,v_{k}}=\{w\in V\mid u>w,\ \mu(u)\ge
\mu(w)+\mu(v_1)+\dots+\mu(v_k)\}.
$$
Then, $M_{u,v_{1},\ldots,v_{k}}$ has maximal elements, and all of them
are children of $u$ and different from $v_1,\dots,v_k$.
\end{enumerate}
\end{lemma}

\begin{proof}
Let us assume that $\mu(u)=\mu(v_1)+\dots+\mu(v_k)$.  Then $u$ cannot
have other children, since if it has any other child $w$, then
$$
\mu(u)\ge \mu(v_1)+\dots+\mu(v_k)+\mu(w)>\mu(v_1)+\dots+\mu(v_k).
$$

Assume now that $\mu(u)>\mu(v_1)+\dots+\mu(v_k)$.  Then, by Lemma
\ref{lem:comp-mu}.(b), $u$ has other children than
$v_{1},\ldots,v_{k}$.  Let $N'$ be the DAG obtained by removing from
$N$ the arcs $(u,v_1),\dots,(u,v_k)$.  For any node $s\in V$, let
$m'_i(s)$ be the number of different paths $s\pathgr l_i$ in $N'$, and
set $\mu'(s)=(m'_1(s),\dots,m'_n(s))$.  Then,
$\mu'(u)=\mu(u)-(\mu(v_1)+\dots+\mu(v_k))$, because the paths
$u\pathgr l_i$ in $N$ that are not in $N'$ are exactly those whose
first visited vertex is one of $v_1,\dots,v_k$.  Moreover, if $w$ is a
descendant of $u$ in $N$, then $\mu'(w)=\mu(w)$, because no path
$w\pathgr l_i$ in $N$ can possibly contain any arc $(u,v_{i})$ (it
would form a cycle with the path $u\pathgr w$).

Then, we have that $\mu(w)=\mu'(w)$ for every $w\in
M_{u,v_{1},\ldots,v_{k}}$, and thus
$$
M_{u,v_{1},\ldots,v_{k}}=\{w\in V\mid u>w\mbox{ in $N$ and }
\mu'(u)\ge \mu'(w)\}.
$$
Now, it turns out that $w\in M_{u,v_{1},\ldots,v_{k}}$ if, and only
if, there exists a non-trivial path $u\pathgr w$ in $N'$.  Indeed, if
there exists a non-trivial path $u\pathgr w$ in $N'$, then there exists
also the same path in $N$, and hence $u>w$ in $N$, and moreover, by
Lemma~\ref{lemanou}.(a) above, $\mu'(u)\ge \mu'(w)$.  Conversely, let
$w$ be a descendant of $u$ in $N$ and assume that $\mu'(u)\ge
\mu'(w)=\mu(w)$.  If $l_{i}\in T_{L}(w)$, then $m_{i}'(u)\geq
m_{i}(w)=1$.  Take the tree path $w\pathgr l_i$ in $N$, which also
exists in $N'$, and any path $u\pathgr l_{i}$ in $N'$.  By Lemma
\ref{lem:unicity-path} (applied to $N'$), either $u\pathgr l_i$ contains
$w\pathgr l_i$ or vice versa.  But the existence of a non-trivial path
$u\pathgr w$ in $N$ prevents the existence of a path $w\pathgr u$ in $N'$.
Therefore, it is $u\pathgr l_i $ that contains $w\pathgr l_i$ and in
particular there exists a path $u\pathgr w$ also in $N'$.

So,
$$
M_{u,v_{1},\ldots,v_{k}}=\{w\in V\mid u>w \mbox{ in $N'$}\}.
$$
Since at least one child of $u$ has survived in $N'$, the previous lemma implies that
this set has maximal elements and they are children of $u$ in $N'$,
and hence they are also children of $u$ in $N$ and they are different
from $v_{1},\ldots,v_{k}$.  \qed
\end{proof}

As we have already mentioned, Lemma~\ref{lem:Mu} applies to any DAG
(and we make use of this fact in the proof of Lemma \ref{Prop:Muv1vk},
because the DAG $N'$ we consider in it need not satisfy the tree-child
condition, and can have more than one root as well as out-degree 1
tree nodes), but Lemma~\ref{Prop:Muv1vk}.(b) need not be true if $N$
does not satisfy the tree-child condition.  Consider again, for
instance, the tree-sibling phylogenetic network $N$ described in
Figure~\ref{fig:contr1}.  In it, $c$ is a maximal element of
$M_{r,d}=\{x\in V\mid r>x,\ \mu(r)\geq \mu(x)+\mu(d)\}$, but it is not
a child of $r$.

We can prove now our main result.

\begin{theorem}\label{thm-isomorphism}
Let $N,N'$ be tree-child phylogenetic networks.  Then, $N \cong N'$
if, and only if, $\mu(N)=\mu(N')$.
\end{theorem}

\begin{proof}
Let $N=(V,E)$ be a tree-child phylogenetic network labeled in $\LL$,
and let $\mu(N)$ be its $\mu$-representation.  Let $V_{\mu}\subset
\NN^n\times \NN$ be the set consisting of the vectors of the form
$(x,i)$ with $x\in \mu(N)$ and $i$ between 1 and the multiplicity of
$x$ in $\mu(N)$.  Consider on $V_{\mu}$ the partial order $\succeq$
defined by
$$
\begin{array}{rl}
(x,i)\succ  (y,j)\Longleftrightarrow & x> y\mbox{ with respect to the
product partial order,}\\ & \mbox{or }x=y\mbox{ and }
i<j.
\end{array}
$$

We know from Lemma~\ref{lemanou} that if $x\in \NN^n$ belongs to
$\mu(N)$ with multiplicity $m\geq 1$, then there exist $m$ nodes in
$N$ with $\mu$-vector $x$, and that they form an elementary path.  For
every node $v\in V$, let $i_{v}$ be the position of $v$ in the
elementary path formed by all nodes with the same $\mu$-vector as $v$.
In particular, if $\mu(v)$ appears in $\mu(N)$ with multiplicity 1,
then $i_v=1$

Lemma~\ref{lemanou} implies then that the mapping
$$
\begin{array}{rcl}
\bar{\mu}: V & \to & V_{\mu}\\
v& \mapsto & \bar{\mu}(v)=(\mu(v),i_{v})
\end{array}
$$
is an isomorphism of partially ordered sets between $V$ with the path
ordering and $V_{\mu}$ with the partial order $\succeq$. Indeed, if $u>v$ then either $\mu(u)>\mu(v)$  or $\mu(u)=\mu(v)$ and $u$ appears before $v$ in the elementary path of all nodes with this $\mu$-vector, and hence
 $i_u<i_v$. Conversely, if 
$\mu(u)>\mu(v)$ or if $\mu(u)=\mu(v)$ and $i_u<i_v$, then
 there exists a non-trivial path $u\pathgr v$.

Therefore, we can
rephrase the last two lemmas as follows:
\begin{enumerate}[(1)]
\item For every $u\in V$ internal, the set
$$
\overline{M}_{u}=\{\bar{\mu}(w)\in \bar{\mu}(V)\mid \bar{\mu}(u)\succ \bar{\mu}(w)\}
$$
has maximal elements, and all of them are images under $\bar{\mu}$ of children of $u$.

\item For every $u\in V$ and for every $v_1,\dots,v_k\in \child(u)$,
\begin{enumerate}[(a)]
\item If $\mu(u)>\mu(v_1)+\dots+v(v_k)$, then the set 
$$
{}\!\overline{M}_{u,v_{1},\ldots,v_{k}}\!\!=\!\{\bar{\mu}(w)\in \bar{\mu}(V)\mid \bar{\mu}(u)\succ \bar{\mu}(w),\
\mu(u)\ge
    \mu(w)+\mu(v_1)+\dots+\mu(v_k)\} $$
    has maximal elements, and all of them are images under $\bar{\mu}$ of children of $u$ other than
    $v_1,\dots,v_k$.
    
\item If $\mu(u)=\mu(v_1)+\dots+\mu(v_k)$, then $u$ has no other
children.

\end{enumerate}
\end{enumerate}

We shall prove that we can recover the set $E$ of arcs in $N$ from
$\mu(N)$.  To do that, consider the set $E_{\mu}\subseteq V_{\mu}\times
V_{\mu}$
obtained through the application of Algorithm~\ref{alg1}:

\begin{algorithm}
\label{alg1}
Given  $\mu(N)$, compute $E_{\mu}$.
\upshape
\begin{tabbing}
\quad \=\quad \=\quad \=\quad \=\quad \=\quad \=\quad \kill
\textbf{begin}\\
\> set $E_{\mu}=\emptyset$\\
\> sort $V_{\mu}$ decreasingly on  the partial order $\succeq$\\
\> \textbf{for each} $(x,i) \in V_{\mu}$ \textbf{do}\\
\> \> set $m=x$\\
\> \> \textbf{while} $m>0$ \textbf{do}\\
\> \>\> \textbf{for each} $(y,j)\in V_{\mu}$ such that $(x,i)\succ (y,j)$ \textbf{do}\\
\> \> \> \> \textbf{if}  $m\geq y$ \textbf{then}\\
\> \> \> \> \> add the arc  $((x,i),(y,j))$ to $E_{\mu}$\\
\> \> \> \> \> set $m=m-y$\\
\textbf{end}
\end{tabbing}
\end{algorithm}

Let us prove that $(u,v)\in E$ if, and only if,
$(\bar{\mu}(u),\bar{\mu}(v))\in E_{\mu}$.  To do that, let $u\in V$ be
an arbitrary node of $N$.  If there is no arc in $E_{\mu}$ with source
$\bar{\mu}(u)$, it can only be because there is no $v\in V$ such that
${\mu}(u)>{\mu}(v)$, and hence $u$ is a leaf of $N$ and $E$
does not contain any arc with source $u$, either.  Otherwise, let
$(\bar{\mu}(u) ,\bar{\mu}(v_1)),(\bar{\mu}(u),\bar{\mu}(v_2)),\ldots,
(\bar{\mu}(u),\bar{\mu}(v_l))$ be the arcs contained in $E_{\mu}$ with
source $\bar{\mu}(u)$, given in the order they are added to $E_{\mu}$.
This entails that $\bar{\mu}(v_1)$ is a maximal element of
$$
\{\bar{\mu}(v)\in \bar{\mu}(V)\mid \bar{\mu}(u)\succ \bar{\mu}(v)\}= \overline{M}_u
$$
and, for each $i=2,\ldots, l$, $\bar{\mu}(v_i)$ is a maximal element of
$$
\begin{array}{l}
\{\bar{\mu}(v)\in \bar{\mu}(V)\mid \bar{\mu}(u) \succ\bar{\mu}(v),
\mu(u)-(\mu(v_1)+\ldots+\mu(v_{i-1}))\geq \mu(v)\}\\
\qquad =\! \{v\in V\mid \bar{\mu}(u) \succ\bar{\mu}(v),\ \mu(u)\geq
\mu(v)+\mu(v_1)+\ldots+\mu(v_{i-1})\}\!=\! \overline{M}_{u,v_1,\ldots,v_{i-1}}.
\end{array}
$$
Therefore, as we have recalled in points (a) and (b.1) above, the
nodes $v_{1},\ldots,v_{l}$ are children of $u$ in $N$, that is,
$(u,v_1),(u,v_2),\ldots,(u,v_l)\in E$.  On the other hand, the
algorithm adds arcs $(\bar{\mu}(u),\bar{\mu}(v_{i}))$ to $E_{\mu}$
until it happens either that $\mu(u)=\mu(v_{1})+\cdots+\mu(v_{l})$, in
which case, by (b.2), $v_{1},\ldots,v_{l}$ are exactly the children of
$u$ in $N$, or that the set of nodes is exhausted and
$\mu(u)>\mu(v_{1})+\cdots+\mu(v_{l})$: but the latter cannot happen,
because $\mu(u)$ must be the sum of the $\mu$-vectors of its children
in $N$.  Thus, in summary, $\child(u)=\{v_{1},\ldots,v_{l}\}$ in $N$
and thus $(u,v_1),(u,v_2),\ldots,(u,v_l)$ are also all the arcs
contained in $E$ with source $u$.
 
This proves that $E=\{(u,v)\mid \bar{\mu}(u),\bar{\mu}(v))\in E_{\mu}\}$, as we claimed.  Now, if $N=(V,E)$ and $N'=(V',E')$ are two
tree-child phylogenetic networks such that $\mu(N)=\mu(N')$, we have
bijections
$$
V\longleftrightarrow V_{\mu}=V'_{\mu}\longleftrightarrow V'.
$$
Renaming in $V$ the nodes of $N'$ through this bijection
$V'\longrightarrow V$, we obtain a phylogenetic network $N''=(V,E'')$
isomorphic to $N'$ and such that $\mu(N'')=\mu(N')=\mu(N)$.  Let
$E_{\mu}$ and $E''_{\mu}$ be, respectively, the sets or arcs obtained
by applying the previous algorithm to $N$ and $N''$.  Since
$\mu(N)=\mu(N'')$, we have that $E_{\mu}=E_{\mu}''$, and hence
 $$
 E=\{(u,v)\mid \bar{\mu}(u),\bar{\mu}(v))\in E_{\mu}\}=\{(u,v)\mid \bar{\mu}(u),\bar{\mu}(v))\in E_{\mu}''\}=E''.
 $$
 This implies that $N=N''$ and therefore
$N\cong N'$ as DAGs.  Now, this isomorphism clearly preserves the
leaves' labels, because it preserves $\mu$-vectors.  Therefore, $N\cong N'$
also as $S$-DAGs.

This proves the ``if'' implication in the statement.  Of course, the
``only if'' implication is obvious.  \qed
\end{proof}

To recover, up to isomorphism, a tree-child phylogenetic network $N$
from its $\mu$-representation $\mu(N)$, it is enough to compute the
set $V_\mu$ associated to the multiset $\mu(N)$, then to apply
Algorithm~\ref{alg1} to compute the set of arcs $E_\mu$, and finally
to label each leaf of the resulting DAG, which will have the form
$(\delta_i^{(n)},m_i)$ with $m_i$ the multiplicity of $\delta_i^{(n)}$ in $\mu(N)$, with the corresponding label $l_i$.


\begin{example}
Let us apply this procedure to the $\mu$-representation of the
tree-child phylogenetic network $N$ depicted in
Fig.~\ref{fig:exemple1}.  From the multiset $\mu(N)$ described in
Table~\ref{tab:taula1}, we obtain the following set $V_{\mu}$, which
we give sorted decreasingly on $\succ$ (and, to simplify the
description of the application of the algorithm, we give names $x_i$
to its elements):
$$
\begin{array}{rl}
V_{\mu}= & \{x_{1}=((1,2,3,4,1),1),\
x_{2}=((0,1,2,3,1),1),\
x_{3}=((0,1,1,2,1),1),\\ & \ \,
x_{4}=((0,1,1,1,1),1),\
x_{5}=((1,1,1,1,0),1),\
x_{6}=((0,1,1,1,0),1),\\ & \ \,
x_{7}=((0,1,1,1,0),2),\
x_{8}=((0,0,1,1,0),1),\
x_{9}=((0,0,1,1,0),2),\\ & \ \,
x_{10}=((1,0,0,0,0),1),\
x_{11}= ((0,1,0,0,0),1),\
x_{12}= ((0,0,1,0,0),1),\\ & \ \,
x_{13}=((0,0,0,1,0),1),\
x_{14}=((0,0,0,1,0),2),\
x_{15}=((0,0,0,0,1),1)\}.
\end{array}
$$
We shall denote the first and the second component of each $x_{k}$ by 
$\mu_{k}$ and $i_{k}$, respectively.

We begin with an empty set of arcs:
$$
E_{\mu}  =\emptyset.
$$
Then we proceed with the \textbf{for each} in Algorithm
\ref{alg1}, visiting all elements of $V_{\mu}$ in the given order:
\begin{itemize}
\item[($x_1$)] We set $m=(1,2,3,4,1)$.  Then, since $m\geq  \mu_{2}$, we add
$(x_{1},x_{2})$ to $E_{\mu}$ and we set $m=m-\mu_{2}=(1,1,1,1,0)$.  The
first next element $x_{k}$ of $V_{\mu}$ with $\mu_{k}\leq m$ is $x_{5}$.  Then,
we add $(x_{1},x_{5})$ to $E_{\mu}$ and we set
$m=m-\mu_{5}=(0,0,0,0,0)$.  This makes us to stop with $x_{1}$.  At the
end of this step, we have
$$
E_{\mu}=\{(x_{1},x_{2}),(x_{1},x_{5})\}.
$$

\item[($x_{2}$)] We set $m=(0,1,2,3,1)$.  Since $m\geq  \mu_{3}$, we
add $(x_{2},x_{3})$ to $E_{\mu}$ and we set
$m=m-\mu_{3}=(0,0,1,1,0)$.  The first next element $x_{k}$  of $V_{\mu}$
with $\mu_{k}\leq m$ is $x_{8}$.  Then, we
add $(x_{2},x_{8})$ to $E_{\mu}$, we set $m=m-\mu_{8}=(0,0,0,0,0)$, and
we stop.  So, at the end of this step, we have
$$
E_{\mu}=\{(x_{1},x_{2}),(x_{1},x_{5}),(x_{2},x_{3}),(x_{2},x_{8})\},
$$

\item[($x_{3}$)] We set $m=(0,1,1,2,1)$.  Then, since $m\geq  \mu_{4}$, we
add $(x_{3},x_{4})$ to $E_{\mu}$ and we set $m=m-\mu_{4}=(0,0,0,1,0)$.
The first next element $x_{k}$ of $V_{\mu}$ with $\mu_{k}\leq m$ is $x_{13}$.  Then, we
add $(x_{3},x_{13})$ to $E_{\mu}$, we set
$m=m-\mu_{13}=(0,0,0,0,0)$, and we stop.  At the end of this step, we have
$$
E_{\mu}=\{(x_{1},x_{2}),(x_{1},x_{5}),(x_{2},x_{3}),(x_{2},x_{8}),(x_{3},x_{4}),
(x_{3},x_{13})\}.
$$

\item[($x_{4}$)] We set $m=(0,1,1,1,1)$.  The first element $x_{k}$ with
$\mu_{k}\leq m$ is $x_{6}$, and therefore we add $(x_{4},x_{6})$ to
$E_{\mu}$ and we set $m=m-\mu_{6}=(0,0,0,0,1)$.  The first  next element $x_{k}$ of
$V_{\mu}$ with $\mu_{k}\leq m$ is $x_{15}$.  Then, we add $(x_{4},x_{15})$
to $E_{\mu}$, we set $m=m-\mu_{15}=(0,0,0,0,0)$, and we stop.  At the
end of this step, we have
$$
E_{\mu}=\{(x_{1},x_{2}),(x_{1},x_{5}),(x_{2},x_{3}),(x_{2},x_{8}),(x_{3},x_{4}),
(x_{3},x_{13}), (x_{4},x_{6}), (x_{4},x_{15})\}.
$$

\item[($x_{5}$)] We set $m=(1,1,1,1,0)$.  Since $m\geq \mu_{6}$, we add
$(x_{5},x_{6})$ to $E_{\mu}$ and we set $m=m-\mu_{6}=(1,0,0,0,0)$.  The
first next element $x_{k}$ of $V_{\mu}$ with $\mu_{k}\leq m$ is
$x_{10}$.  Then, we add
$(x_{5},x_{10})$ to $E_{\mu}$.  Now $m=m-x_{10}=(0,0,0,0,0)$, and we
stop.  At the end of this step, we have
$$
\begin{array}{rl}
E_{\mu}=& \{(x_{1},x_{2}),(x_{1},x_{5}),(x_{2},x_{3}),(x_{2},x_{8}),(x_{3},x_{4}),
(x_{3},x_{13}), (x_{4},x_{6}), (x_{4},x_{15}),\\
& \ \, (x_{5},x_{6}),(x_{5},x_{10})\}.
\end{array}
$$

\item[($x_{6}$)] We set $m=(0,1,1,1,0)$. 
Since $m\geq \mu_{7}$, we add $(x_{6},x_{7})$ to $E_{\mu}$ and we set
$m=m-\mu_{7}=(0,0,0,0,0)$, and we stop.
At the end of this step, we have
$$
\begin{array}{rl}
E_{\mu}=& \{(x_{1},x_{2}),(x_{1},x_{5}),(x_{2},x_{3}),(x_{2},x_{8}),(x_{3},x_{4}),
(x_{3},x_{13}), (x_{4},x_{6}), (x_{4},x_{15}),\\
& \ \, (x_{5},x_{6}),(x_{5},x_{10}),(x_{6},x_{7})\}.
\end{array}
$$

\item[($x_{7}$)] We set (again) $m=(0,1,1,1,0)$. 
Since $m\geq \mu_{8}$, we add
$(x_{7},x_{8})$ to $E_{\mu}$ and we set $m=m-\mu_{8}=(0,1,0,0,0)$.  The
first next element $x_{k}$ of $V_{\mu}$ with $\mu_{k}\leq m$ is
$x_{11}$.  Then, we add
$(x_{7},x_{11})$ to $E_{\mu}$.  Now $m=m-\mu_{11}=(0,0,0,0,0)$, and we
stop.  At the end of this step, we have
$$
\begin{array}{rl}
E_{\mu}=& \{(x_{1},x_{2}),(x_{1},x_{5}),(x_{2},x_{3}),(x_{2},x_{8}),(x_{3},x_{4}),
(x_{3},x_{13}), (x_{4},x_{6}), (x_{4},x_{15}),\\
& \ \, (x_{5},x_{6}),(x_{5},x_{10}),(x_{6},x_{7}),(x_{7},x_{8}),
(x_{7},x_{11}) \}.
\end{array}
$$

\item[($x_{8}$)] We set $m=(0,0,1,1,0)$. Since $m\geq \mu_{9}$, we
add $(x_{8},x_{9})$ to $E_{\mu}$, we set $m=m-\mu_{9}=(0,0,0,0,0)$, and
we stop.  At the end of this step, we have
$$
\begin{array}{rl}
E_{\mu}=& \{(x_{1},x_{2}),(x_{1},x_{5}),(x_{2},x_{3}),(x_{2},x_{8}),(x_{3},x_{4}),
(x_{3},x_{13}), (x_{4},x_{6}), (x_{4},x_{15}),\\
& \ \,
(x_{5},x_{6}),(x_{5},x_{10}),(x_{6},x_{7}),(x_{7},x_{8}),(x_{7},x_{11}),
(x_{8},x_{9}) \}.
\end{array}
$$

\item[($x_{9}$)] We set $m=(0,0,1,1,0)$.  The first element $x_{k}$ in
$V_{\mu}$ with $\mu_{k}\leq m$ is $x_{12}$, and then we add
$(x_{9},x_{12})$ to $E_{\mu}$ and we set $m=m-\mu_{12}=(0,0,0,1,0)$.
The first next element $x_{k}$ in
$V_{\mu}$ with $\mu_{k}\leq m$ is
$x_{13}$.  Then, we
add $(x_{9},x_{13})$ to $E_{\mu}$.  Now $m=m-\mu_{13}=(0,0,0,0,0)$, and
we stop.  At the end of this step, we have
$$
\begin{array}{rl}
E_{\mu}\!\!=\!\!& \{(x_{1},x_{2}),(x_{1},x_{5}),(x_{2},x_{3}),(x_{2},x_{8}),(x_{3},x_{4}),
(x_{3},x_{13}), (x_{4},x_{6}), (x_{4},x_{15}),\\
& \ \,
(x_{5},x_{6}),(x_{5},x_{10}),(x_{6},x_{7}),(x_{7},x_{8}),(x_{7},x_{11}),
(x_{8},x_{9}), (x_{9},x_{12}), (x_{9},x_{13}) \}.
\end{array}
$$

\item[$(x_{10}$)] Since $m=(1,0,0,0,0)$ has only one non-zero entry,
and $x_{10}$ is the only element of $V_{\mu}$ with $\mu_{k}=m$,
the algorithm does not find any arc in $E_{\mu}$ with head $x_{10}$.  The
same happens with $x_{11}$, $x_{12}$, and $x_{15}$.

\item[$(x_{13}$)] We set $m=(0,0,0,1,0)$. Since $m\geq \mu_{14}$, we
add $(x_{13},x_{14})$ to $E_{\mu}$, we set $m=m-\mu_{14}=(0,0,0,0,0)$, and
we stop.  At the end of this step, we have
$$
\begin{array}{rl}
E_{\mu}\!\!=\!\! & \{(x_{1},x_{2}),(x_{1},x_{5}),(x_{2},x_{3}),(x_{2},x_{8}),(x_{3},x_{4}),
(x_{3},x_{13}), (x_{4},x_{6}), (x_{4},x_{15}),\\
& \ \,
(x_{5},x_{6}),(x_{5},x_{10}),(x_{6},x_{7}),(x_{7},x_{8}),(x_{7},x_{11}),
(x_{8},x_{9}), (x_{9},x_{12}), (x_{9},x_{13}) ,\\
& \ \,
(x_{13},x_{14})\}.
\end{array}
$$

\item[$(x_{14}$)] Since $m=(0,0,0,1,0)$ has only one non-zero entry,
and no other element of $V_{\mu}$ after $x_{14}$ has this first
component, 
the algorithm does not find any arc in $E_{\mu}$ with head $x_{14}$.

\end{itemize}

\begin{figure}[htb]
\begin{center}
    \begin{tikzpicture}[thick,>=stealth]
  \draw(1,0) node[tre] (x_1) {}; \etq{x_1}
  \draw(2,-1) node[tre] (x_2) {}; \etq{x_2}
  \draw(3,-2) node[tre] (x_3) {}; \etq{x_3}
  \draw(4,-3) node[tre] (x_4) {}; \etq{x_4}
  \draw(0,-1) node[tre] (x_5) {}; \etq{x_5}
  \draw(1,-3) node[hyb] (x_6) {}; \etq{x_6}
  \draw(1,-4) node[tre] (x_7) {}; \etq{x_7}
  \draw(2,-4) node[hyb] (x_8) {}; \etq{x_8}
  \draw(2,-5) node[tre] (x_9) {}; \etq{x_9}
  \draw(3,-5) node[hyb] (x_{13}) {}; \etq{x_{13}}
  \draw(0,-6) node[tre] (x_{10}) {}; \etq{x_{10}}
  \draw(1,-6) node[tre] (x_{11}) {}; \etq{x_{11}}
  \draw(2,-6) node[tre] (x_{12}) {}; \etq{x_{12}}
  \draw(3,-6) node[tre] (x_{14}) {}; \etq{x_{14}}
  \draw(4,-6) node[tre] (x_{15}) {}; \etq{x_{15}}
  \draw[->](x_1)--(x_2);
  \draw[->](x_1)--(x_5);
  \draw[->](x_2)--(x_8);
  \draw[->](x_2)--(x_3);
  \draw[->](x_5)--(x_{10});
  \draw[->](x_5)--(x_6);
  \draw[->](x_3)--(x_4);
  \draw[->](x_3)--(x_{13});
  \draw[->](x_4)--(x_6);
  \draw[->](x_4)--(x_{15});
  \draw[->](x_6)--(x_7);
  \draw[->](x_7)--(x_8);
  \draw[->](x_7)--(x_{11});
  \draw[->](x_8)--(x_9);
  \draw[->](x_9)--(x_{12});
  \draw[->](x_9)--(x_{13});
  \draw[->](x_{13})--(x_{14});
  \end{tikzpicture}
\end{center}
\caption{\label{fig:recons2} The DAG recovered from the
$\mu$-representation of the phylogenetic network in
Fig.~\ref{fig:exemple1}.}
\end{figure}
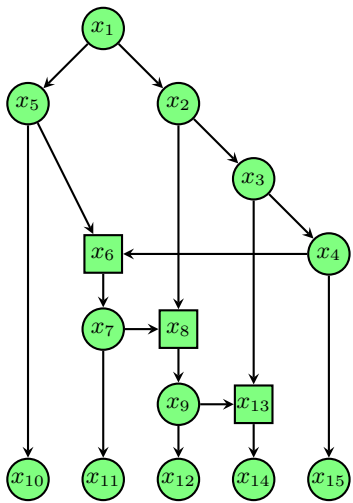

The DAG $(V_{\mu},E_{\mu})$ obtained up to now is depicted in
Fig.~\ref{fig:recons2}.
Finally, we would label $1$, $2$, $3$, $4$, and $5$ the nodes
$x_{10}$, $x_{11}$, $x_{12}$, $x_{14}$, and $x_{15}$, respectively.
The resulting DAG labeled in $\{1,\ldots,5\}$ is clearly isomorphic to
the tree-child phylogenetic network $N$ in Fig.~\ref{fig:exemple1}.
\end{example}

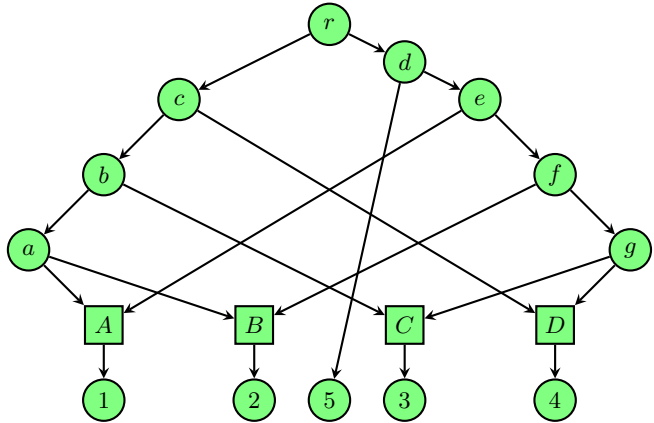
\begin{figure}[htb]
\begin{center}
    \begin{tikzpicture}[thick,>=stealth]
    \draw(0,0) node[tre] (r) {}; \etq r
    \draw(-2,-1) node[tre] (c) {};\etq c
    \draw(-3,-2) node[tre] (b) {};\etq b
    \draw(-4,-3) node[tre] (a) {};\etq a
    \draw(2,-1) node[tre] (e) {};\etq e
    \draw(3,-2) node[tre] (f) {};\etq f
    \draw(4,-3) node[tre] (g) {};\etq g
    \draw(-3,-4) node[hyb] (A) {};\etq A
    \draw(-1,-4) node[hyb] (B) {};\etq B
    \draw(1,-4) node[hyb] (C) {};\etq C
    \draw(3,-4) node[hyb] (D) {};\etq D
    \draw(-3,-5) node[tre] (1) {};\etq 1
    \draw(-1,-5) node[tre] (2) {};\etq 2
    \draw(1,-5) node[tre] (3) {};\etq 3
    \draw(3,-5) node[tre] (4) {};\etq 4
    \draw(1,-0.5) node[tre] (d) {};\etq d
    \draw(0,-5) node[tre] (5) {};\etq 5
    \draw[->](r)--(d);
    \draw[->](r)--(c);
    \draw[->](d)--(e);
    \draw[->](d)--(5);
    \draw[->](c)--(b);
    \draw[->](c)--(D);
    \draw[->](e)--(A);
    \draw[->](e)--(f);
    \draw[->](b)--(a);
    \draw[->](b)--(C);
    \draw[->](f)--(g);
    \draw[->](f)--(B);
    \draw[->](a)--(A);
    \draw[->](a)--(B);
    \draw[->](g)--(C);
    \draw[->](g)--(D);
    \draw[->](A)--(1);
    \draw[->](B)--(2);
    \draw[->](C)--(3);
    \draw[->](D)--(4);
  \end{tikzpicture}
\end{center}
\caption{\label{fig:contr3}This phylogenetic network has the same
$\mu$-representation as the one in Fig.~\ref{fig:contr1}.}
\end{figure}

\begin{remark}
The thesis of Theorem~\ref{thm-isomorphism} need not hold if $N_1$ and
$N_2$ do not satisfy the tree-child condition.  Indeed, it is not
difficult to check that the tree-sibling phylogenetic network given in
Figure~\ref{fig:contr3} has the same $\mu$-representation as the one
given in Figure~\ref{fig:contr1}, but they are not isomorphic as
$\LL$-DAGs.
\end{remark}

\section{The $\mu$-distance for tree-child phylogenetic networks}
\label{sec:mu-distance}

For every pair of DAGs $N_1$ and $N_2$ labeled in the same set $S$,
let
$$
d_\mu(N_1,N_2)=|\mu(N_1)\bigtriangleup \mu(N_2)|,
$$
where the symmetric difference $\bigtriangleup$ refers to multisets:
if a vector belongs to $\mu(N_1)$ with multiplicity $a$ and to
$\mu(N_2)$ with multiplicity $b$, then it belongs to
$\mu(N_1)\bigtriangleup \mu(N_2)$ with multiplicity $|a-b|$, and hence
it contributes $|a-b|$ to $|\mu(N_1)\bigtriangleup \mu(N_2)|$.

\begin{theorem}\label{th:mudist}
Let $N_{1},N_{2},N_{3}$ be tree-child phylogenetic networks on the
same set of taxa.  Then:
\begin{enumerate}[(a)]
\item $d_\mu(N_1,N_2)\ge 0$

\item $d_\mu(N_1,N_2)= 0$ if, and only if, $N_1\cong N_2$

\item $d_{\mu}(N_{1},N_{2})=d_{\mu}(N_{2},N_{1})$
  
\item $d_\mu(N_1,N_3)\leq d_\mu(N_1,N_2)+d_\mu(N_2,N_3)$.
\end{enumerate}
\end{theorem}

\begin{proof}
(a), (c), and (d) are direct consequences of the properties of the
symmetric difference, and (b) is a  consequence of
Theorem~\ref{thm-isomorphism}.\qed
\end{proof}

Therefore, $d_{\mu}$ defines a distance on the class of all tree-child
phylogenetic networks: we shall call it the \emph{$\mu$-distance}.

We have shown in Sect.~\ref{sec:mu-representation} that the $\mu$-representation of an $S$-DAG can be computed in polynomial time. Now, given two $S$-DAGs $N_1=(V_1,E_1)$ and $N_2=(V_2,E_2)$ with $n$ leaves together with their $\mu$-representations $\mu(N_1)$ and $\mu(N_2)$, the simple Algorithm~\ref{alg00} performs a simultaneous traversal of the internal nodes of $N_1$ and $N_2$, sorted by their $\mu$-vectors, in order to compute the $\mu$-distance $d_\mu(N_1,N_2)$ in $O(n|V|)$ time, where $|V|=\max(|V_1|,|V_2|)$.

\begin{algorithm}
\label{alg00}
Given the $\mu$-representations $\mu(N_1)$ and $\mu(N_2)$ of two $S$-DAGs $N_1=(V_1,E_1)$ and $N_2=(V_2,E_2)$, compute $d_\mu(N_1,N_2)$.
\upshape
\begin{tabbing}
\quad \=\quad \=\quad \=\quad \=\quad \=\quad \=\quad \kill
\textbf{begin} \\
\> let $V_L$ be the set of leaves, common to $N_1$ and $N_2$ \\
\> sort $X_1 = V_1 \setminus (V_1)_L$ increasingly according to the lex ordering of the $\mu$-vectors \\
\> sort $X_2 = V_2 \setminus (V_2)_L$ increasingly according to the lex ordering of the $\mu$-vectors \\
\> set $d = 0$ \\
\> \textbf{while} $X_1 \neq \emptyset$ and $X_2 \neq \emptyset$ \textbf{do} \\
\> \> let $x_1$ and $x_2$ be the first element of $X_1$ and $X_2$, respectively \\
\> \> \textbf{case} $\mu(x_1) < \mu(x_2)$ \\
\> \> \> set $X_1 = X_1 \setminus \{x_1\}$ \\
\> \> \> set $d = d + 1$ \\
\> \> \textbf{case} $\mu(x_1) > \mu(x_2)$ \\
\> \> \> set $X_2 = X_2 \setminus \{x_2\}$ \\
\> \> \> set $d = d + 1$ \\
\> \> \textbf{otherwise} \\
\> \> \> set $X_1 = X_1 \setminus \{x_1\}$ \\
\> \> \> set $X_2 = X_2 \setminus \{x_2\}$ \\
\> \textbf{return} $d + |X_1| + |X_2|$ \\
\textbf{end}
\end{tabbing}
\end{algorithm}


\begin{example}\label{ex:computmudist0}
Consider the tree $T$ and the network $N$ in Fig.~\ref{fig:treemetric}.
Their $\mu$-representa\-tions are
$$
\begin{array}{rl}
\mu(T)& =\{(1,0,0),(0,1,0),(0,0,1),(0,1,1),(1,1,1)\}\\
\mu(N) & =\{(1,0,0),(0,1,0),(0,0,1), (0,1,0),
(0,1,1),(0,2,1),(1,2,1)\}
\end{array}
$$
and therefore
$$
\mu(T)\bigtriangleup\mu(N)=\{(0,1,0),(1,1,1),(0,2,1),(1,2,1)\}
$$
from which we obtain $d_{\mu}(T,N)=4$.
\end{example}

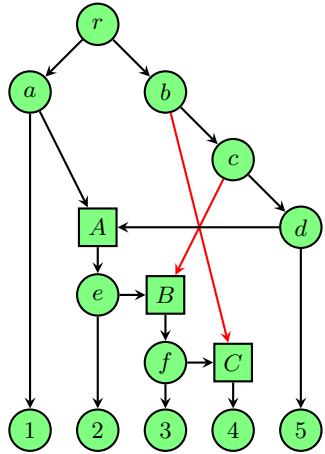
\begin{figure}[htb]
\begin{center}
    \begin{tikzpicture}[thick,>=stealth,scale=0.9]
\draw (3,7) node[tre] (r) {}; \etq r
\draw (2,6) node[tre] (a) {}; \etq a
\draw (4,6) node[tre] (b) {}; \etq b
\draw (2,1) node[tre] (1) {}; \etq 1
\draw (5,5) node[tre] (c) {}; \etq c
\draw (3,4) node[hyb] (A) {}; \etq A
\draw (6,4) node[tre] (d) {}; \etq d
\draw (3,3) node[tre] (e) {}; \etq e
\draw (4,3) node[hyb] (B) {}; \etq B
\draw (6,1) node[tre] (5) {}; \etq 5
\draw (3,1) node[tre] (2) {}; \etq 2
\draw (4,2) node[tre] (f) {}; \etq f
\draw (5,2) node[hyb] (C) {}; \etq C
\draw (4,1) node[tre] (3) {}; \etq 3
\draw (5,1) node[tre] (4) {}; \etq 4
\draw [->](r)--(a);
\draw [->](r)--(b);
\draw [->](a)--(1);
\draw [->](a)--(A);
\draw [red,->](b)--(C);
\draw [->](b)--(c);
\draw [red,->](c)--(B);
\draw [->](c)--(d);
\draw [->](A)--(e);
\draw [->](d)--(A);
\draw [->](d)--(5);
\draw [->](e)--(2);
\draw [->](e)--(B);
\draw [->](B)--(f);
\draw [->](f)--(3);
\draw [->](f)--(C);
\draw [->](C)--(4);
\end{tikzpicture}
\end{center}
\caption{\label{fig:exemple1bis}
The tree-child phylogenetic network $N'$ compared in
Example~\ref{ex:computmudist} with the tree-child phylogenetic network
$N$ from Fig.~\ref{fig:exemple1}.}
\end{figure}

\begin{example}\label{ex:computmudist}
It was shown in \cite{cardona.ea:07a} that the tree-child phylogenetic
network $N$ depicted in Fig.~\ref{fig:exemple1} could not be
distinguished from the tree-child phylogenetic network $N'$ depicted in
Fig.~\ref{fig:exemple1bis} below using the tripartition metric.  We
have already given in Table~\ref{tab:taula1} the $\mu$-vectors of the
nodes of $N$.  In Table~\ref{tab:taula1bis} we give the $\mu$-vectors
of the nodes of $N'$.  From these tables we get that
$$
|\mu(N)\bigtriangleup\mu(N')|=\big\{(0,1,1,2,1),(0,1,2,2,1)\big\},
$$
which implies that $d_{\mu}(N,N')=2$.
\end{example}

\begin{table}[htb]
   \centering
\begin{tabular}{|c|c|c||c|c|c||c|c|c|}
\hline
node & height & $\mu$-vector  & node  & height & $\mu$-vector & node  & height & $\mu$-vector  \\
\hline
$1$ & 0 &{}\   $(1,0,0,0,0)$\ {}  &
$C$ & 1 &{}\   $(0,0,0,1,0)$\ {}  &
$a$ & 6 &{}\  $(1,1,1,1,0)$\ {} \\
\hline
$2$ & 0 & $(0,1,0,0,0)$ &
$f$ & 2 & {}\ $(0,0,1,1,0)$\ {}  &
$d$ & 6 & $(0,1,1,1,1)$\\
\hline
$3$ & 0 & $(0,0,1,0,0)$ &
$B$ & 3 & $(0,0,1,1,0)$ &
$c$ & 7 & $(0,1,2,2,1)$\\
\hline
$4$ & 0 & $(0,0,0,1,0)$ &
$e$ & 4 & $(0,1,1,1,0)$ &
$b$ & 8 & $(0,1,2,3,1)$\\
\hline
$5$ & 0 & $(0,0,0,0,1)$ &
 $A$ & 5 & $(0,1,1,1,0)$ &
$r$ & 9 & $(1,2,3,4,1)$\\
\hline
 \end{tabular} 
 \smallskip
\caption{$\mu$-vectors of the nodes of the network depicted in
Fig.~\ref{fig:exemple1bis}.}
   \label{tab:taula1bis}
\end{table} 

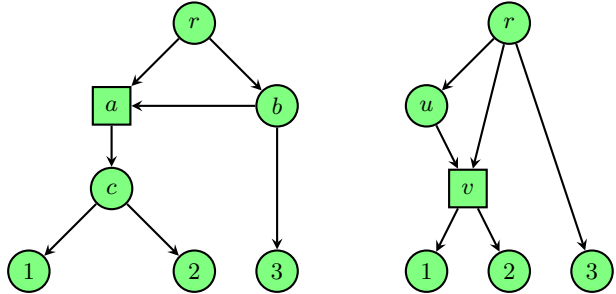
\begin{figure}[htb]
\begin{center}
	  \begin{tikzpicture}[thick,>=stealth,scale=1.1]
	    \draw(0,0) node[tre] (r) {}; \etq r
	    \draw(-1,-1) node[hyb] (a) {}; \etq a
	    \draw(-1,-2) node[tre] (c) {}; \etq c
	    \draw(1,-1) node[tre] (b) {}; \etq b
	    \draw(-2,-3) node[tre] (1) {}; \etq 1
	    \draw(0,-3) node[tre] (2) {}; \etq 2
	    \draw(1,-3) node[tre] (3) {}; \etq 3      
	  \draw[->] (r)--(a);
	  \draw[->] (r)--(b);
	  \draw[->] (a)--(c);
	\draw[->] (b)--(a);

	  \draw[->] (c)--(1);
	  \draw[->] (c)--(2);
	  \draw[->] (b)--(3);
	  \end{tikzpicture}
  \qquad\qquad
  \begin{tikzpicture}[thick,>=stealth,scale=1.1]
      \draw(0,0) node[tre] (r) {}; \etq r
        \draw(-1,-1) node[tre] (u) {}; \etq u
        \draw(-0.5,-2) node[hyb] (v) {}; \etq v
         \draw(-1,-3) node[tre] (1) {}; \etq 1
        \draw(0,-3) node[tre] (2) {}; \etq 2
        \draw(1,-3) node[tre] (3) {}; \etq 3
        \draw[->] (r)--(u);
        \draw[->] (r)--(v);
        \draw[->] (r)--(3);
        \draw[->] (u)--(v);
        \draw[->] (v)--(1);
        \draw[->] (v)--(2);
  \end{tikzpicture}
\end{center}
\caption{\label{fig:mudist1}
Two tree-child phylogenetic networks $N$ (left) and $N'$ (right) at
$\mu$-distance 1.}
\end{figure}

\begin{example}
Let $N$ be any tree-child phylogenetic network labeled in $S$ and let
$v$ be any internal node of it.  Let $N'$ be a tree-child phylogenetic
network obtained by adding to $N$ a new internal node $v'$, an arc
$(v,v')$, and then distributing the children of $v$ between $v$ and
$v'$ so that $N'$ remains tree-child and $v'$ does not become a leaf.
Then $\mu(N)\bigtriangleup \mu(N')=\{\mu(v')\}$ and therefore
$d_{\mu}(N,N')=1$.

So, expanding a node into an arc yields $\mu$-distance 1, just as it
happens with Robinson-Foulds distance for phylogenetic trees.  This is
consistent with the fact, which we shall prove later, that the
$\mu$-distance extends  the Robinson-Foulds
distance to tree-child networks: cf.~Theorem~\ref{thm:mudist=RF}
below.  But, contrary to the tree case, two tree-child phylogenetic
networks can be at $\mu$-distance 1 without any one of them being
obtained by expanding a node into an arc in the other one.  Consider,
for instance, the tree-child phylogenetic networks $N$ and $N'$
labeled in $\{1,2,3\}$ depicted in Fig.~\ref{fig:mudist1}.  Their
$\mu$-representations are
$$
\begin{array}{rl}
\mu(N) &
=\big\{(1,0,0),(0,1,0),(0,0,1),(1,1,0),(1,1,0),(1,1,1),(2,2,1)\big\}\\
\mu(N') & 
=\big\{(1,0,0),(0,1,0),(0,0,1),(1,1,0),(1,1,0),(2,2,1)\big\}
\end{array}
$$
and thus $d_{\mu}(N,N')=1$.

It should also be noticed that, also against what happens in the tree
case, collapsing an arc in a tree-child phylogenetic network $N$ into
a node (that is, given an arc $(v,v')$, removing $v'$ and this arc,
and replacing every other arc with tail or head $v'$ by a new arc with
tail or head, respectively, $v$) need not produce a network at
$\mu$-distance 1 of $N$: for instance, if $v$ and $v'$ hybridize in
$N$, or if $v'$ is a non-strict descendant of $v$.  We leave to the
interested reader to draw specific counterexamples.
\end{example}

\begin{figure}
\centering
\includegraphics[width=0.48\linewidth]{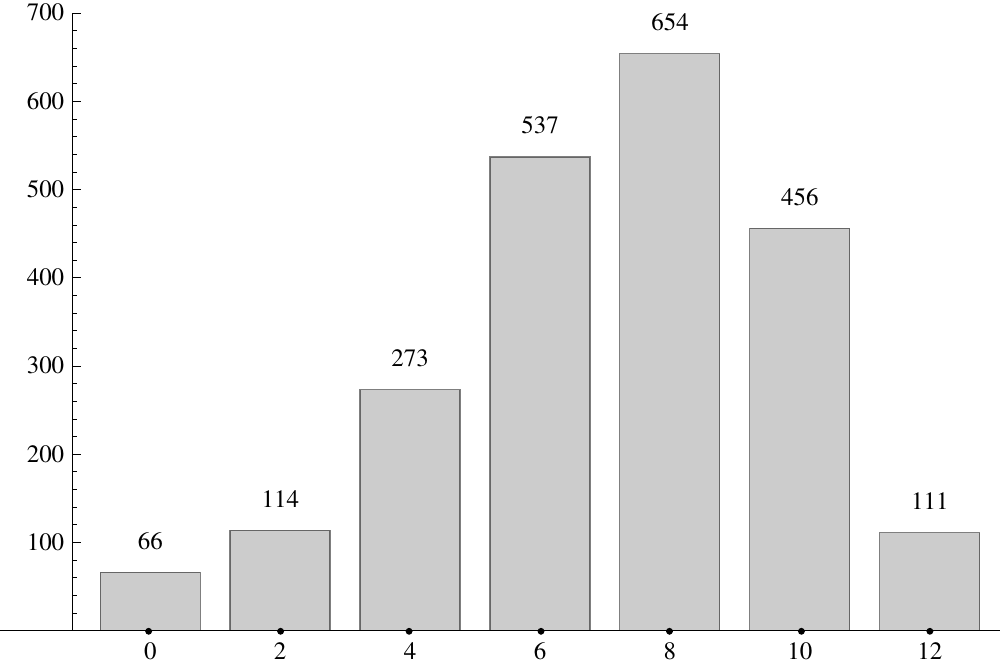}\quad
\includegraphics[width=0.48\linewidth]{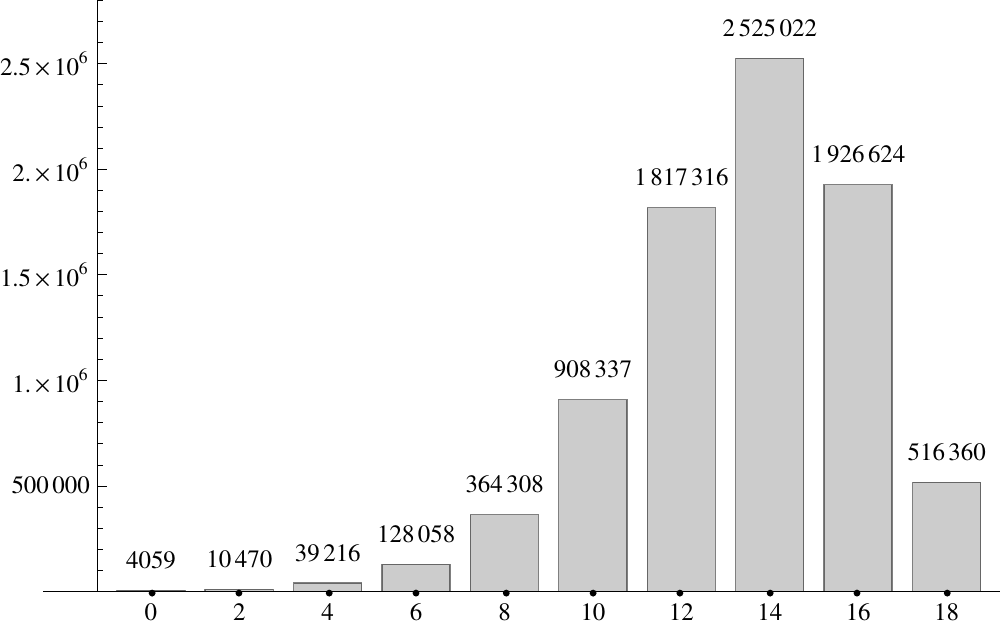}
\caption{Histograms of $\mu$-distances between unordered pairs of binary tree-child
phylogenetic trees with 3 (left) and 4 (right) leaves.} 
    \label{fig:hists}
\end{figure}

\begin{example}\label{ex:3leaves}
There exist 66 pairwise non-isomorphic binary tree-child phylogenetic
networks with 3 leaves.  All of them have an even number of internal
nodes, and therefore the $\mu$-distance between two of them is always
an even number.  In Proposition \ref{prop:distbound} below we shall
see that this $\mu$-distance is smaller than or equal to 12.  The
left-hand side histogram in Fig.~\ref{fig:hists} shows the
distribution of distances between unordered pairs of such networks.

In a similar way, there exist 4059 pairwise non-isomorphic binary
tree-child phylogenetic networks with 4 leaves.  Again, all of them have
an even number of internal nodes, and therefore the $\mu$-distance
between two of them is always even, and in Proposition
\ref{prop:distbound} we shall see that it is smaller than or equal to
18.  The right-hand side histogram in Fig.~\ref{fig:hists} shows the
corresponding distribution of distances.

See the Supplementary Material for more details.
\end{example}

Every phylogenetic tree is a tree-child phylogenetic network, and, as
we have already mentioned, it turns out that the restriction of this
$\mu$-distance to the class of phylogenetic trees is the
Robinson-Foulds metric.

\begin{theorem}\label{thm:mudist=RF}
For every phylogenetic trees $T_{1},T_{2}$ on the same set of taxa
$S$,
$$
d_{\mu}(T_1,T_2)=d_{RF}(T_1,T_2).
$$
\end{theorem}

\begin{proof}
Let $S=\{l_{1},\ldots,l_{n}\}$.  The uniqueness of paths in trees
implies that, if $T=(V,E)$ is any phylogenetic tree labeled in $S$,
then, for every $u\in V$,
$$
m_{i}(u)=\left\{
\begin{array}{ll}
1 & \mbox{ if $l_{i}\in C_{L}(u)$}\\
0 & \mbox{ if $l_{i}\notin C_{L}(u)$}
\end{array}
\right.
$$
Therefore, the $\mu$-vector of a node $u$ of a phylogenetic tree
labeled in $S$ is the image of its cluster $C_{L}(u)$ under the
bijection between the powerset $\mathcal{P}(S)$ of $S$ and $\{0,1\}^n$
that sends each subset $A$ of $S$ to its characteristic vector
$$
\chi_{A}=(\chi_{A}(l_1),\ldots,\chi_{A}(l_n)),
\mbox{ with }
\chi_{A}(l_i)=\left\{
\begin{array}{ll}
1 & \mbox{ if $l_{i}\in A$}\\
0 & \mbox{ if $l_{i}\notin A$}
\end{array}
\right.
$$

Then, given two phylogenetic trees $T_{1}$ and $T_{2}$ on the set of
taxa $S$,  this bijection $\mathcal{P}(S)\longrightarrow \{0,1\}^n$ transforms
the sets $C_{L}(T_{1})$ and $C_{L}(T_{2})$ of clusters of their nodes into
their $\mu$-representations $\mu(T_{1})$ and $\mu(T_{2})$,
respectively, and hence the symmetric difference of the former into
the symmetric difference of the latter. Therefore,
$$
|\pi(T_{1})\bigtriangleup \pi(T_{2})|=|C_{L}(T_{1})\bigtriangleup C_{L}(T_{2})|=
|\mu(T_{1})\bigtriangleup \mu(T_{2})|,
$$
as we claimed. \qed
\end{proof}

The $\mu$-distance $d_\mu$ takes integer values.  Its smallest
non-zero value is 1 but it can be arbitrarily large.  If we bound the
in-degree of the hybrid nodes of the networks, then we can compute the
diameter of the resulting subclass of phylogenetic networks.

Let $\mathcal{TCN}^{n,m}$ be henceforth  the class of all tree-child
phylogenetic networks on a fixed set of taxa of $n$ elements, without
out-degree 1 tree nodes and with all their hybrid nodes of in-degree
at most $m$.

\begin{proposition}
\label{prop:distbound}
For every $N_{1},N_{2}\in \mathcal{TCN}^{n,m}$,
$$
d_{\mu}(N_{1},N_{2})\leq 2(m+1)(n-1)
$$
and there exist pairs of networks in $\mathcal{TCN}^{n,m}$ at $\mu$-distance $2(m+1)(n-1)$.
\end{proposition}

\begin{proof}
Let $N_{1},N_{2}\in \mathcal{TCN}^{n,m}$.  By Proposition
\ref{lem:bound}.(c), each one of $\mu(N_1), \mu(N_2)$ has at most
$(m+2)(n-1)+1$ elements, from which at least the $n$ $\mu$-vectors
corresponding to the leaves will appear in both sets.  Therefore
    $$
    d_{\mu}(N_1,N_2)\leq 2\big((m+2)(n-1)+1\big)-2n=2(m+1)(n-1).
    $$

To find a pair of networks in $\mathcal{TCN}^{n,m}$ at distance $2(m+1)(n-1)$,
 let $N$ be the tree-child phylogenetic
network with $n$ leaves and $n-1$ hybrid nodes of in-degree $m$
described in Example \ref{ex:bignetwork}. A simple argument by
induction shows that
$$
\begin{array}{rl}
\mu(h_{n}) & = (0,\ldots,0,0,0,0,1)\\
\mu(v_{n-1,k}) & =(0,\ldots,0,0,0,1,k)\mbox{ for every $k=1,\ldots,m$}\\
\mu(h_{n-1}) & =(0,\ldots,0,0,0,1,m)\\
\mu(v_{n-2,k}) & =(0,\ldots,0,0,1,k,km)\mbox{ for every $k=1,\ldots,m$}\\
\mu(h_{n-2}) & =(0,\ldots,0,0,1,m,m^2)\\
\mu(v_{n-3,k}) & =(0,\ldots,0,1,k,km,km^2)\mbox{ for every
$k=1,\ldots,m$}\\
\ldots
\end{array}
$$
and, in general,
$$
\begin{array}{rl}
\mu(h_{n-j}) &\! =\!(\overbrace{0,\ldots,0}^{n-j-1},1,m,m^2,\ldots,m^j)\mbox{ for every
$j=0,\ldots,n-2$}\\
\mu(v_{n-j,k}) &\! =\!(\overbrace{0,\ldots,0}^{n-j-2},1,k,km,\ldots,km^j)\mbox{ for every
$k=1,\ldots,m$,\ $j=1,\ldots,n-1$}
\end{array}
$$
Therefore, $\mu(N)$ contains, beside the $\mu$-vectors
$\delta_{i}^{(n)}$ of leaves, all vectors of the form
$$
(\overbrace{0,\ldots,0}^{n-j-2},1,k,km,\ldots,km^j),\qquad
k=1,\ldots,m-1,\ j=1,\ldots,n-1
$$
with multiplicity 1, and all vectors of the form
$$
(\overbrace{0,\ldots,0}^{n-j-1},1,m,m^2,\ldots,m^j),\qquad
j=1,\ldots,n-2
$$
with multiplicity 2.

Now let $N'$ be the tree-child phylogenetic network in
$\mathcal{TCN}^{n,m}$ obtained by performing the same construction
starting with the binary phylogenetic tree described by the Newick
string
$$
(n,(n-1,(n-2,\ldots,(2,1)\ldots ))).
$$
The same argument shows that $\mu(N')$ contains, again beside the $\mu$-vectors
$\delta_{i}^{(n)}$ of leaves, all vectors of the form
$$
(km^j,\ldots,km,k,1,\overbrace{0,\ldots,0}^{n-j-2}),\qquad
k=1,\ldots,m-1,\ j=1,\ldots,n-1
$$
with multiplicity 1, and all vectors of the form
$$
(m^j,\ldots,m^2,m,1,\overbrace{0,\ldots,0}^{n-j-1}),\qquad
j=1,\ldots,n-2
$$
with multiplicity 2.

Then, $\mu(N)$ and $\mu(N')$ have no $\mu$-vector of internal node in 
common, and since each one has $(m+1)(n-1)$ internal nodes, this implies
that $d_{\mu}(N,N')= 2(m+1)(n-1)$.\qed
\end{proof}

This result allows us to normalize the $\mu$-distance on
$\mathcal{TCN}^{n,m}$.

\begin{corollary}
\label{cor:distnorm}
The mapping
$$
\begin{array}{rcl}
d'_{\mu}: \mathcal{TCN}^{n,m}\times \mathcal{TCN}^{n,m} & \to & \RR\\
(N_1 ,N_2 )\quad & \mapsto & \dfrac{1}{2(m+1)(n-1)}d_{\mu}(N_1,N_2)
\end{array}
$$
is a distance $\mathcal{TCN}^{n,m}$ that takes values in the unit
interval $[0,1]$.
\end{corollary}

\section{The alignment of tree-child phylogenetic networks}
\label{sec:mu-align}

Let $N_{1}=(V_{1},E_{1})$ and $N_{2}=(V_{2},E_{2})$ be two tree-child
phylogenetic networks labeled in the same set
$\LL=\{l_{1},\ldots,l_{n}\}$.  For simplicity, we assume that they
don't have out-degree 1 tree nodes, and therefore, if two nodes  in one of these networks  have the same $\mu$-vector, then they must be a hybrid node and  its only child.

For every $v_{1}\in V_{1}$ and
$v_{2}\in V_{2}$, let
$$
\begin{array}{rl}
H(v_{1},v_{2}) & =\displaystyle
\sum_{i=1}^{n}|m_{i}(v_{1})-m_{i}(v_{2})|
\\
\chi(v_{1},v_{2}) & =\left\{
\begin{array}{ll}
0 & \mbox{ if $v_{1},v_{2}$ are of the same type (both tree nodes or
both hybrid)}\\
1 & \mbox{ if $v_{1},v_{2}$ are of different type}
\end{array}
\right.
\end{array}
$$
Notice that $H(v_{1},v_{2}) $ is the Manhattan, or $L_1$,
distance between $\mu(v_{1})$ and $\mu(v_{2})$. The advantage of this distance over the Euclidean distance is that it takes integer values on $\NN^n$.

Define finally the \emph{weight} of the pair $(v_{1},v_{2})$ as
$$
w(v_{1},v_{2})=H(v_{1},v_{2})+\frac{\chi(v_{1},v_{2})}{2n}.
$$

To fix ideas, assume that $|V_{1}|\leq |V_{2}|$.  Then, given a
\emph{matching between $N_{1}$ and $N_{2}$}, that is, an injective
mapping $M:V_{1}\to V_{2}$ that preserves leaves and their labels, its
\emph{total weight} is defined as
$$
w(M)=\sum_{v\in V_{1}} w(v,M(v)).
$$
An \emph{optimal alignment} between $N_{1}$ and $N_{2}$ is a matching
with the smallest total weight.  Such an optimal alignment can be
computed in time $O\big((|V_{1}|+|V_{2}|)^3\big)$ using the Hungarian
algorithm \cite{kuhn:55,munkres:57}.

\begin{proposition}
A matching $M$ between $N_{1}$ and $N_{2}$ is an optimal alignment if, and
only if, it minimizes the sum
$$
\sum_{v\in V_{1}\setminus V_{L}} H(v,M(v))
$$
and, among those matchings minimizing this sum, it maximizes the number
of nodes that are sent to nodes of the same type.
\end{proposition}

\begin{proof}
Let $M:V_{1}\to V_{2}$ be any matching. Then
$$
w(M)=\sum_{v\in V_{1}} w(v,M(v))=\sum_{v\in V_{1}} H(v,M(v))+
\frac{1}{2n}\sum_{v\in V_{1}}\chi(v,M(v)).
$$
The first addend is a positive integer, while the second addend is
strictly smaller than 1, because by Proposition \ref{lem:bound}.(a)
both $N_{1}$ and $N_{2}$ have at most $n-1$ hybrid nodes, and
therefore $\sum_{v\in V_{1}}\chi(v,M(v))\leq 2(n-1)$.  Therefore,
$\sum_{v\in V_{1}} H(v,M(v))$ is the integer part of $w(M)$.  This
implies that $w(M)\leq w(M') $ if, and only if,
$$
\sum_{v\in V_{1}} H(v,M(v))\leq \sum_{v\in V_{1}} H(v,M'(v))
$$
and, if the latter are equal, also
$$
\frac{1}{2n}\sum_{v\in V_{1}}\chi(v,M(v))
\leq
\frac{1}{2n}\sum_{v\in V_{1}}\chi(v,M'(v)),
$$
from where the statement clearly follows.
\qed
\end{proof}

\begin{remark}
If we restrict this alignment method to phylogenetic trees, 
the weight
of a pair of nodes $(v_1,v_2)$ is simply $|C_L(v_1)\bigtriangleup
C_L(v_2)|$.  This can be seen as an unnormalized version of the score
used in \emph{TreeJuxtaposer} \cite{munzner.ea:03}.
\end{remark}

\begin{remark}
Let $N=(V,E)$ and $N'=(V',E')$ be two tree-child phylogenetic networks
without out-degree 1 tree nodes.  If they are isomorphic, the
isomorphism between them is an optimal alignment of total weight 0.
The converse implication is clearly false in general: a matching of
total weight 0 need not be an isomorphism.  Consider for instance the
optimal alignment between the phylogenetic trees described by the
Newick strings $(1,2,3)$ and $(1,(2,3))$.

But if there exists an alignment $M$ (obviously optimal) between $N$
and $N'$ of total weight 0 and if $|V|=|V'|$, then $M$ is a bijection
between $V$ and $V'$ that preserves the $\mu$-vectors, because
$$
\sum_{v\in V}|\mu(v)-\mu(M(v))|\leq w(M)=0\Rightarrow
\mu(v)=\mu(M(v))\mbox{ for every $v\in V$,}
$$
and therefore  $\mu(N)=\mu(N')$, which implies, by
Theorem \ref{thm-isomorphism}, $N\cong N'$.
\end{remark}


Given two $S$-DAGs $N_1=(V_1,E_1)$ and $N_2=(V_2,E_2)$ together with their $\mu$-representations $\mu(N_1)$ and $\mu(N_2)$, the simple Algorithm~\ref{alg000} computes the total weight $w(M)$ of an optimal alignment $M$ between $N_1$ and $N_2$ in $O((|V_1|+|V_2|)^3)$ time.

\begin{algorithm}
\label{alg000}
Given the $\mu$-representations $\mu(N_1)$ and $\mu(N_2)$ of two $S$-DAGs $N_1=(V_1,E_1)$ and $N_2=(V_2,E_2)$, compute the total weight of an optimal alignment between $N_1$ and $N_2$.
\upshape
\begin{tabbing}
\quad \=\quad \=\quad \=\quad \=\quad \=\quad \=\quad \kill
\textbf{begin} \\
\> let $G=((V_1\setminus (V_1)_L) \sqcup V_2\setminus (V_2)_L,E)$ be a complete bipartite graph with $|V_1|+|V_2|-2n$ vertices \\
\> \textbf{for each} $x_1 \in V_1\setminus (V_1)_L$ and $x_2 \in V_2\setminus (V_2)_L$ \textbf{do} \\
\> \> set $\mathop{weight}[x_1,x_2] = \mathop{abs}(\mu(x_1)-\mu(x_2))$ \\
\> \> \textbf{if} $x_1$ and $x_2$ are not both tree nodes or both hybrid \textbf{then} \\
\> \> \> add $1/2n$ to $\mathop{weight}[x_1,x_2]$ \\
\> let $M:V_1\setminus (V_1)_L\to V_2\setminus (V_2)_L$ be a minimum-weight bipartite matching of $G$ \\
\> extend $M$ by sending each $v\in (V_1)_L$ to the leaf of $N_2$ with the same label\\
\> \textbf{return} $M$\\
\> \textbf{return} $\sum_{i=1}^{|V_1|} \mathop{weight}[v_1,M(v_1)]$ \\
\textbf{end}
\end{tabbing}
\end{algorithm}

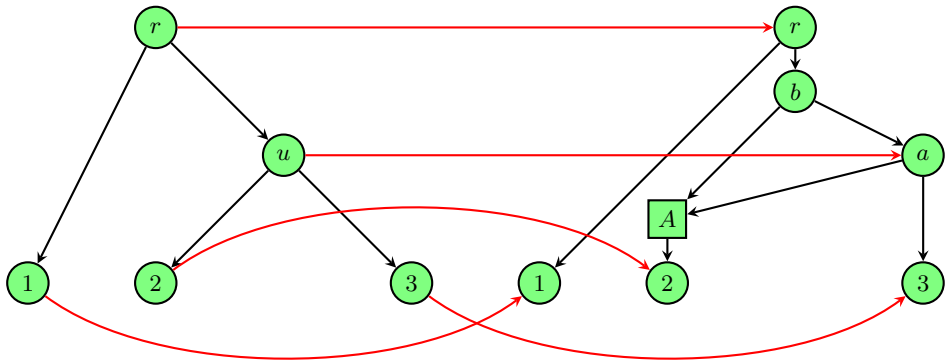
\begin{figure}[htb]
\begin{center}
\begin{tikzpicture}[thick,>=stealth,scale=1.7]
	    \draw(0,0) node[tre] (r) {}; \etq r
	    \draw(1,-1) node[tre] (u) {}; \etq u
	    \draw(-1,-2) node[tre] (1) {}; \etq 1
	    \draw(0,-2) node[tre] (2) {}; \etq 2
	    \draw(2,-2) node[tre] (3) {}; \etq 3   
	    \draw(5,0) node[tre] (rb) {};\draw (rb) node {$r$};
	\draw(5,-0.5) node[tre] (b) {}; \etq b
	\draw(6,-1) node[tre] (a) {}; \etq a
	\draw(4,-1.5) node[hyb] (A) {}; \etq A
	 \draw(4,-2) node[tre] (2b) {}; \draw (2b) node {$2$};
	\draw(6,-2) node[tre] (3b) {}; \draw (3b) node {$3$};
	\draw(3,-2) node[tre] (1b) {}; \draw (1b) node {$1$};
	  \draw[->] (r)--(u);
	  \draw[->] (r)--(1);
	  \draw[->] (u)--(3);
	  \draw[->] (u)--(2);
        \draw[->] (rb)--(b);
        \draw[->] (rb)--(1b);
        \draw[->] (b)--(a);
        \draw[->] (b)--(A);
        \draw[->] (a)--(A);
        \draw[->] (A)--(2b);
        \draw[->] (a)--(3b);    
         \draw[->,color=red] (u) -- (a);
    \draw[->,color=red] (r) -- (rb);
     \draw[->,color=red] (1) .. controls (0,-2.75) and (2,-2.75) ..(1b);
          \draw[->,color=red] (2) .. controls (1,-1.25) and (3,-1.25) ..(2b);
     \draw[->,color=red] (3) .. controls (3,-2.75) and (5,-2.75) ..(3b);

  \end{tikzpicture}
\end{center}
\caption{\label{fig:align1} 
An optimal alignment  between the tree and the
galled tree in Fig.~\ref{fig:treemetric}.}
\end{figure}

\begin{example}\label{ex:align1}
Consider the tree $T$ and the galled tree $N$ depicted in
Fig.~\ref{fig:treemetric}.  The total weight of the matching between
$T$ and $N$ that sends the root of $T$ to the root of $N$ and the node
$u$ of $T$ to the tree node $a$ of $N$ is 
$$
w(r,r)+w(u,a)+w(1,1)+w(2,2)+w(3,3)=0+1+0+0+0=1,
$$
and hence, since $\mu(T)\not\subseteq \mu(N)$, it is an optimal
alignment: see Fig.~\ref{fig:align1}.
\end{example}

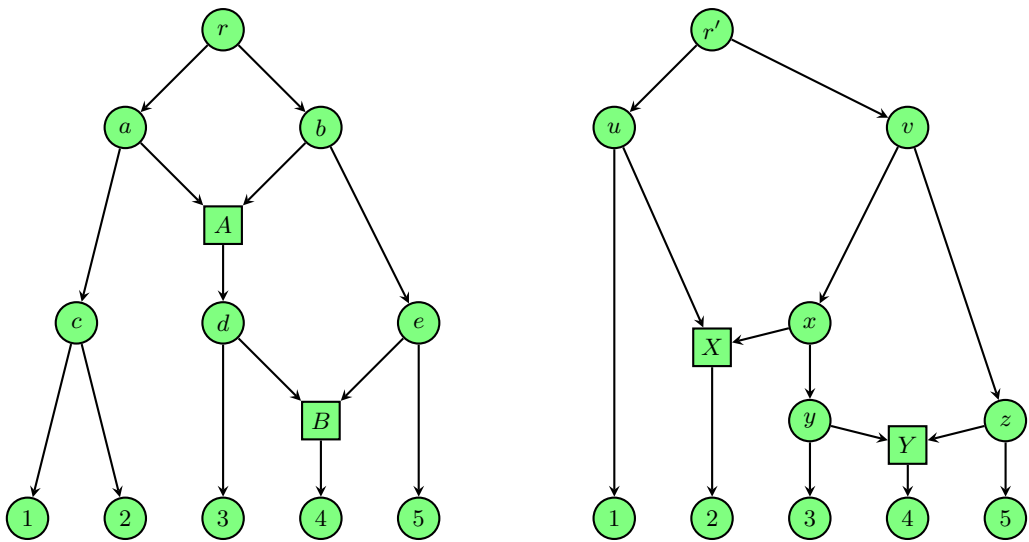
\begin{figure}[htb]
\begin{center}
\begin{tikzpicture}[thick,>=stealth,scale=1.3]
\draw(0,0) node[tre] (r) {}; \etq r
\draw(-1,-1) node[tre] (a) {}; \etq a
\draw(1,-1) node[tre] (b) {}; \etq b
\draw(0,-2) node[hyb] (A) {}; \etq A
\draw(-1.5,-3) node[tre] (c) {}; \etq c
\draw(0,-3) node[tre] (d) {}; \etq d
\draw(2,-3) node[tre] (e) {}; \etq e
\draw(1,-4) node[hyb] (B) {}; \etq B
\draw(-2,-5) node[tre] (1) {}; \etq 1
\draw(-1,-5) node[tre] (2) {}; \etq 2
\draw(0,-5) node[tre] (3) {}; \etq 3
\draw(1,-5) node[tre] (4) {}; \etq 4
\draw(2,-5) node[tre] (5) {}; \etq 5
\draw(5,0) node[tre] (rb) {};\draw (rb) node {$r'$};
\draw(4,-1) node[tre] (u) {}; \etq u
\draw(7,-1) node[tre] (v) {}; \etq v
\draw(6,-3) node[tre] (x) {}; \etq x
\draw(5,-3.25) node[hyb] (X) {}; \etq X
\draw(6,-4) node[tre] (y) {}; \etq y
\draw(8,-4) node[tre] (z) {}; \etq z
\draw(7,-4.25) node[hyb] (Y) {}; \etq Y
\draw(4,-5) node[tre] (1b) {}; \draw (1b) node {$1$};
\draw(5,-5) node[tre] (2b) {}; \draw (2b) node {$2$};
\draw(6,-5) node[tre] (3b) {}; \draw (3b) node {$3$};
\draw(7,-5) node[tre] (4b) {}; \draw (4b) node {$4$};
\draw(8,-5) node[tre] (5b) {}; \draw (5b) node {$5$};
\draw[->] (r)--(a);
\draw[->] (r)--(b);
\draw[->] (a)--(A);
\draw[->] (b)--(A); 
\draw[->] (a)--(c); 
\draw[->] (b)--(e);
\draw[->] (A)--(d);
\draw[->] (c)--(1);
\draw[->] (c)--(2);
\draw[->] (d)--(3);
\draw[->] (d)--(B);
\draw[->] (e)--(B);
\draw[->] (B)--(4);
\draw[->] (e)--(5);
\draw[->] (rb)--(u);
\draw[->] (rb)--(v);
\draw[->] (u)--(X);
\draw[->] (v)--(x);
\draw[->] (x)--(X);
\draw[->] (x)--(y);
\draw[->] (y)--(Y);
\draw[->] (v)--(z);
\draw[->] (z)--(Y);
\draw[->] (u)--(1b);
\draw[->] (X)--(2b);
\draw[->] (y)--(3b);
\draw[->] (Y)--(4b);
\draw[->] (z)--(5b);
 \end{tikzpicture}
\end{center}
\caption{\label{fig:align2a} 
The networks $N$ (left) and $N'$ (right) used in Example~\ref{ex:align2}.}
\end{figure}

\begin{figure}[htb]
\begin{center}
\begin{tikzpicture}[thick,>=stealth,scale=1.3]
\draw(0,0) node[tre] (r) {}; \etq r
\draw(-1,-1.5) node[tre] (a) {}; \etq a
\draw(1,-0.75) node[tre] (b) {}; \etq b
\draw(0,-3.25) node[hyb] (A) {}; \etq A
\draw(-1.5,-2.5) node[tre] (c) {}; \etq c
\draw(0,-4) node[tre] (d) {}; \etq d
\draw(2,-4.5) node[tre] (e) {}; \etq e
\draw(1,-5) node[hyb] (B) {}; \etq B
\draw(-2,-5.75) node[tre] (1) {}; \etq 1
\draw(-1,-5.75) node[tre] (2) {}; \etq 2
\draw(0,-5.75) node[tre] (3) {}; \etq 3
\draw(1,-5.75) node[tre] (4) {}; \etq 4
\draw(2,-5.75) node[tre] (5) {}; \etq 5
\draw(5,0) node[tre] (rb) {};\draw (rb) node {$r'$};
\draw(4,-2.5) node[tre] (u) {}; \etq u
\draw(7,-0.75) node[tre] (v) {}; \etq v
\draw(6,-1.5) node[tre] (x) {}; \etq x
\draw(5,-3.25) node[hyb] (X) {}; \etq X
\draw(6,-4) node[tre] (y) {}; \etq y
\draw(8,-4.5) node[tre] (z) {}; \etq z
\draw(7,-5) node[hyb] (Y) {}; \etq Y
\draw(4,-5.75) node[tre] (1b) {}; \draw (1b) node {$1$};
\draw(5,-5.75) node[tre] (2b) {}; \draw (2b) node {$2$};
\draw(6,-5.75) node[tre] (3b) {}; \draw (3b) node {$3$};
\draw(7,-5.75) node[tre] (4b) {}; \draw (4b) node {$4$};
\draw(8,-5.75) node[tre] (5b) {}; \draw (5b) node {$5$};
\draw[->] (r)--(a);
\draw[->] (r)--(b);
\draw[->] (a)--(A);
\draw[->] (b)--(A); 
\draw[->] (a)--(c); 
\draw[->] (b)--(e);
\draw[->] (A)--(d);
\draw[->] (c)--(1);
\draw[->] (c)--(2);
\draw[->] (d)--(3);
\draw[->] (d)--(B);
\draw[->] (e)--(B);
\draw[->] (B)--(4);
\draw[->] (e)--(5);
\draw[->] (rb)--(u);
\draw[->] (rb)--(v);
\draw[->] (u)--(X);
\draw[->] (v)--(x);
\draw[->] (x)--(X);
\draw[->] (x)--(y);
\draw[->] (y)--(Y);
\draw[->] (v)--(z);
\draw[->] (z)--(Y);
\draw[->] (u)--(1b);
\draw[->] (X)--(2b);
\draw[->] (y)--(3b);
\draw[->] (Y)--(4b);
\draw[->] (z)--(5b);
\draw[->,color=red] (r) -- (rb);
\draw[->,color=red] (b) -- (v);
\draw[->,color=red] (a) -- (x);
\draw[->,color=red] (A) -- (X);
\draw[->,color=red] (c) -- (u);
\draw[->,color=red] (d) -- (y);
\draw[->,color=red] (e) -- (z);
\draw[->,color=red] (B) -- (Y);
%
%
%
 \end{tikzpicture}
\end{center}
\caption{\label{fig:align2b} 
An optimal alignment between the networks in Fig.~\ref{fig:align2a}.}
\end{figure}
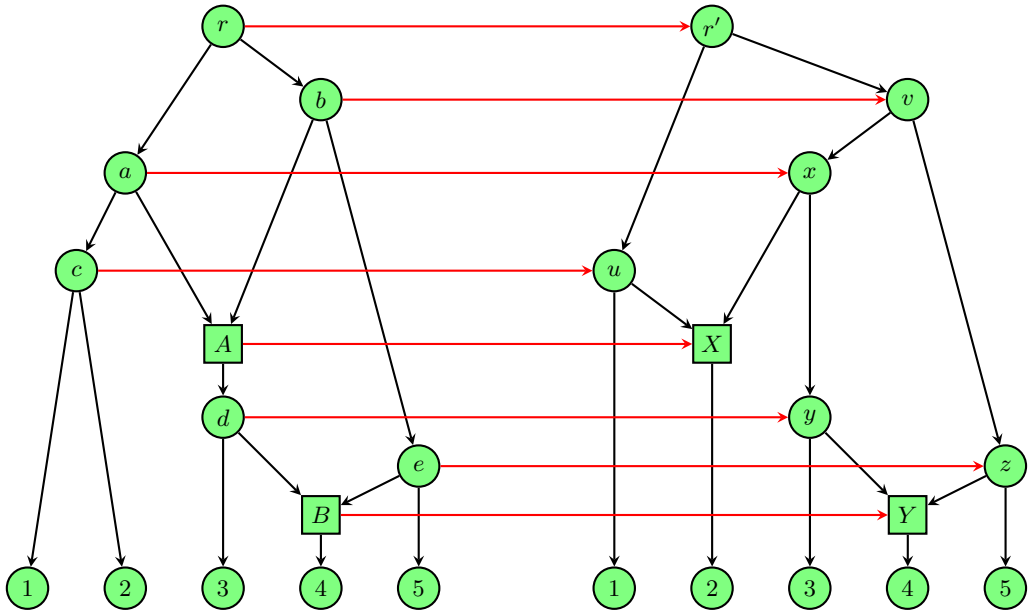

\begin{example}\label{ex:align2}
Consider the tree-child phylogenetic networks $N$ and $N'$ labeled in
$\{1,\ldots,5\}$ given in Fig.~\ref{fig:align2a}.  The $\mu$-vectors
of their internal nodes are given in Table~\ref{tab:taula-align2}.
Table \ref{tab:taula-pesos1} gives the values of $w(x,y)$ for every internal node $x$ of
$N$ and every internal node $y$ of $N'$.  From this table,  the optimal
alignment marked in red in the table and depicted in
Fig.~\ref{fig:align2b} (where, to simplify the picture, the arrows
joining each leaf of $N$ to the homonymous leaf in $N'$   are
omitted) is deduced: its total weight is 8. It is the only optimal
alignment between these networks.
\end{example}

\begin{table}[htb]
   \centering
\begin{tabular}{|c|c||c|c||c|c||c|c|}
\hline
node & $\mu$-vector & node & $\mu$-vector & node & $\mu$-vector & node
& $\mu$-vector \\
\hline
$r$ & $(1,1,2,3,1)$ &
$b$ & $(0,0,1,2,1)$ &
$a$ & $ (1,1,1,1,0)$ &
$A$ & $(0,0,1,1,0)$ \\
\hline
$c$ & $(1,1,0,0,0)$ &
$d$ & $(0,0,1,1,0)$ &
$e$ & $(0,0,0,1,1)$ &
$B$ & $(0,0,0,1,0)$ \\
\hline
\hline
$r'$ & $(1,2,1,2,1)$ &
$u$ & $(1,1,0,0,0)$ &
$v$ & $(0,1,1,2,1)$&
$x$ & $(0,1,1,1,0)$\\
\hline
$y$ & $(0,0,1,1,0)$ &
$z$ & $(0,0,0,1,1)$ &
$X$ & $(0,1,0,0,0)$ &
$Y$ & $(0,0,0,1,0)$ \\
\hline
 \end{tabular} 
 \smallskip
\caption{$\mu$-vectors of the internal nodes of the networks depicted in
Fig.~\ref{fig:align2a}.}
   \label{tab:taula-align2}
\end{table}

\begin{table}[htb]
$$
\begin{array}{|c||c|c|c|c|c|c|c|c|}
\hline
m & r' & u & v & x & y & z & X & Y \\ \hline\hline
r & \red{3} & 6 & 3 & 5 & 6 & 6 & \,7.1\, & \,7.1\, \\ \hline
b & 3 & 6 & \red{1} & 3 & 2 & 2 & 5.1 & 3.1 \\ \hline
a & 3 & 2 & 3 & \red{1} & 2 & 4 & 3.1 & 3.1 \\ \hline
\,A\, & \,5.1\, & \,4.1\, & \,3.1\, & \,1.1\, & \,0.1\, & \,2.1\, & \red{3} & 1 \\ \hline
c & 5 & \red{0} & 5 & 3 & 4 & 4 & 1.1 & 3.1 \\ \hline
d & 5 & 4 & 3 & 1 & \red{0} & 2 & 3.1 & 1.1 \\ \hline
e & 5 & 4 & 3 & 3 & 2 & \red{0} & 3.1 & 1.1 \\ \hline
B & 6.1 & 3.1 & 4.1 & 2.1 & 1.1 & 1.1 & 2 & \red{0} \\ \hline
 \end{array} 
 $$
 \smallskip
\caption{Weights of the pairs of internal nodes of the networks depicted in
Fig.~\ref{fig:align2a}. }
   \label{tab:taula-pesos1}
\end{table}

A web tool that computes an optimal alignment of two tree-child
phylogenetic networks with the same leaves and without out-degree 1
tree nodes is available at the Supplementary Material web page.

\section{Conclusion}

In this paper we have presented two methods for comparing pairs of
tree-child phylogenetic networks: a metric and an alignment algorithm.
While the former quantifies the similarity between two networks in a
way that allows to soundly establish whether a network is more similar
to a second one than to a third one, the latter allows the interactive
visualization of the differences between two networks.  They are
respectively the first true distance and the first alignment method
defined on a meaningful class of phylogenetic networks strictly
extending the class of phylogenetic trees.  Since the class of
tree-child phylogenetic networks includes the galled trees, this
distance and this alignment can be used to compare the latter.

Tree-child phylogenetic networks have been recently proposed by S. J.
Wilson as the class of networks where to look for meaningful
phylogenies, but for the moment no reconstruction algorithm for
tree-child phylogenetic networks has been developed.  So, it remains
an interesting open question to characterize the sets of sequences
whose evolution can be explained by means of a tree-child network and
to provide an algorithm to reconstruct this network, as well as to
characterize the computational complexity of these problems.

On the other hand, several reconstruction methods for time-consistent
tree-sibling phylogenetic networks have been proposed by Nakhleh and
collaborators.  Since no true distance for these networks is known so
far, it is an interesting open question whether our distance and
alignment method can be extended to these networks or not.

\section*{Acknowledgment}

The research described in this paper has been partially supported by
the Spanish CICYT project TIN 2004-07925-C03-01 GRAMMARS, by Spanish
DGI projects MTM2006-07773 COMGRIO and MTM2006-15038-C02-01, and by EU project INTAS IT
04-77-7178.

\section*{Supplementary Material}
The Supplementary Material referenced in the paper  is available at the url\\
\url{http://bioinfo.uib.es/\~{}recerca/phylonetworks/mudistance}.

\end{document}